\documentclass[aps,11pt,prd,longbibliography]{revtex4-1}

\usepackage{feynmp-auto}
\usepackage{amsmath}
\usepackage[utf8]{inputenc}

\usepackage{bbm}
\usepackage{soul}	
\usepackage{xcolor}

\usepackage{epsfig}
\usepackage{color}
\usepackage{slashed}
\usepackage{comment}
\usepackage{epstopdf}
\usepackage{array}
\usepackage{booktabs}
\usepackage{mathrsfs}
\usepackage[euler]{textgreek}
\usepackage{amsmath}
\usepackage{amsthm}
\usepackage{amsfonts}
\usepackage{amssymb}
\usepackage{graphicx}
\graphicspath{ {Figures/} }
\usepackage[font={small}]{caption}
\usepackage{subcaption}
\usepackage{float}
\usepackage{multirow}
\usepackage[export]{adjustbox}
\usepackage[hang, flushmargin,bottom]{footmisc} 
\usepackage{hyperref}
\hypersetup{
     colorlinks   = true,
     citecolor    = blue
}
\usepackage{placeins}
\usepackage{enumitem}
\usepackage{slashed}
\usepackage{bbm}
\usepackage{appendix}

\usepackage{titlesec}
\titleformat{\chapter}[display]
  {\normalfont\LARGE\bfseries}
  {\chaptertitlename\ \thechapter}{5pt}{\LARGE}
  \titlespacing*{\chapter}{0pt}{-20pt}{35pt}
\usepackage{bigstrut}
\usepackage{pst-node}
\usepackage{pstricks}
\usepackage{physics}
\usepackage{mathtools}
\usepackage{setspace}
\usepackage{fancyhdr}
\usepackage[makeroom]{cancel}
\usepackage{feyn}
\newcommand{\be}{\begin{equation}}
\newcommand{\ee}{\end{equation}}
\newcommand{\bes}{\begin{equation*}}
\newcommand{\ees}{\end{equation*}}

\usepackage{hyperref}
\hypersetup{%
  colorlinks = true,
  linkcolor  = black
}
\usepackage{soul}

\newcommand{\beq}{\begin{equation}}
\newcommand{\eeq}{\end{equation}}

\newcommand{\SU}{\,{\rm SU}}
\newcommand{\U}{\,{\rm U}}

\DeclareUnicodeCharacter{2212}{-}

\AtBeginDocument{%
    \newwrite\bibnotes
    \def\bibnotesext{Notes.bib}
    \immediate\openout\bibnotes=\jobname\bibnotesext
    \immediate\write\bibnotes{@CONTROL{REVTEX41Control}}
    \immediate\write\bibnotes{@CONTROL{%
    apsrev41Control,author="08",editor="1",pages="1",title="0",year="1"}}
     \if@filesw
     \immediate\write\@auxout{\string\citation{apsrev41Control}}%
    \fi
}%

\begin{document}

\title{{\Large{\bf{Neutrino-Dark Matter Connections in Gauge Theories}}}}
\author{Pavel Fileviez P\'erez$^{1}$, Clara Murgui$^{2}$, Alexis D. Plascencia$^{1}$}
\affiliation{$^{1}$Physics Department and Center for Education and Research in Cosmology and Astrophysics (CERCA), 
Case Western Reserve University, Rockefeller Bldg. 2076 Adelbert Rd. Cleveland, OH 44106, USA \\
$^{2}$Departamento de F\'isica Te\'orica, IFIC, Universitat de Valencia-CSIC, 
E-46071, Valencia, Spain}
\email{pxf112@case.edu, clara.murgui@ific.uv.es, alexis.plascencia@case.edu}
%
\begin{abstract}
We discuss the connection between the origin of neutrino masses and the properties of dark matter candidates in the context of gauge extensions of the Standard Model. We investigate minimal gauge theories for neutrino masses where the neutrinos are predicted to be Dirac or Majorana fermions. We find that the upper bound on the effective number of relativistic species provides a strong constraint in the scenarios with Dirac neutrinos. In the context of theories where the lepton number is a local gauge symmetry spontaneously broken at the low scale, the existence of dark matter is predicted from the condition of anomaly cancellation. Applying the cosmological bound on the dark matter relic density, we find an upper bound on the symmetry breaking scale in the multi-TeV region. These results imply we could hope to test simple gauge theories for neutrino masses at current or future experiments.
\end{abstract}
\maketitle

\section{INTRODUCTION}
\label{sec:Introduction}
The origin of neutrino masses is one of the most pressing issues in particle physics today.
The Standard Model (SM) of particle physics needs to be modified in order to account for neutrino masses.  Thanks to the ongoing effort of many experimental collaborations, at present we have constraints on the masses and mixing angles of neutrinos, see for example Ref.~\cite{Esteban:2018azc}. However, we still do not know the type of spectrum, whether CP symmetry is broken in the leptonic sector and whether neutrinos are Dirac or Majorana fermions.

The simplest gauge symmetries we can use to understand the origin of neutrino masses are $B\!-\!L$ or $L$, where $B$ 
and $L$ stand for baryon and lepton numbers, respectively. The neutrinos are Majorana fermions when the 
$B\!-\!L$ $({\rm{or}} \ L)$ symmetry is broken in two units, or they can be Dirac fermions when $B\!-\!L$ $({\rm{or}} \ L)$ 
is conserved or broken in units different than two. In both cases, we can hope to test the mechanism for neutrino masses 
only if the $B\!-\!L$ $({\rm{or}} \ L)$ symmetry breaking scale can be reached at current or future colliders.
Unfortunately, in a large class of models for Majorana neutrino masses based on the seesaw mechanism~\cite{Minkowski:1977sc,GellMann:1980vs,Yanagida:1979as,Mohapatra:1979ia}, the canonical seesaw 
scale can be very large, $M_{\rm{seesaw}} \lesssim 10^{14}$ GeV, which makes the mechanism impossible to falsify.

Recently, we have discussed a simple theory for neutrino masses where the seesaw scale is in the multi-TeV region~\cite{FileviezPerez:2018toq}.
In this context, the same $\U(1)_{B-L}$ gauge symmetry that explains the origin of neutrino masses defines the properties 
of a cold dark matter candidate. Using the cosmological constraints on the dark matter relic density, it was found that the seesaw 
scale must be in the multi-TeV region. Therefore, there is hope to test the origin of neutrino masses 
and the seesaw mechanism at colliders. For other studies of gauged $B\!-\!L$ with a dark matter candidate see Refs.~\cite{
Okada:2010wd,
Basak:2013cga,
Kanemura:2014rpa,
Duerr:2015wfa,
Ma:2015mjd,
Wang:2015saa,
Okada:2016gsh,
Kaneta:2016vkq,
DeRomeri:2017oxa,
Okada:2018ktp,
Escudero:2018fwn}, and for scenarios that explore a connection between the origin of neutrino masses and the dark matter candidate see Refs.~\cite{Ma:2006km,
Boehm:2006mi,
Farzan:2010mr,
Batell:2017rol,
Batell:2017cmf,
Blennow:2019fhy}.

In this article, we investigate possible connections between the origin of neutrino masses and the properties of dark matter candidates in simple gauge extensions of the SM
based on local $B\!-\!L$ or $L$ gauge symmetries. We focus on two main scenarios in the context of $B\!-\!L$ theories: (a) Stueckelberg Scenario and (b) Canonical Seesaw Scenario.
In the Stueckelberg scenario, the $B\!-\!L$ symmetry remains unbroken, the neutrinos are Dirac fermions and the gauge boson acquires mass through the Stueckelberg mechanism.
In the Canonical Seesaw scenario the $\U(1)_{B-L}$ symmetry is spontaneously broken in two units using the Higgs mechanism and the neutrinos are Majorana fermions. In both cases the dark matter 
candidate is a vector-like Dirac fermion charged under the $\U(1)_{B-L}$ symmetry. We study the simplest theories with gauged lepton number, $\U(1)_L$ \cite{FileviezPerez:2011pt,Duerr:2013dza,Schwaller:2013hqa,Perez:2014qfa}, where the existence 
of a dark matter candidate is predicted from anomaly cancellation. Models with gauged lepton number have also been studied in Refs.~\cite{Chao:2010mp,
Aranda:2014zta,
Fornal:2017owa,
Chang:2018vdd,
Chang:2018nid}.  In this case, the lepton number is broken in three units and the neutrinos are predicted to be Dirac fermions, 
while the dark matter candidate is a Majorana fermion. 

We investigate the cosmological constraints on the relic dark matter density and show that the upper bound 
on the symmetry breaking scale in those theories is in the multi-TeV scale. Therefore, one could test these theories for neutrino masses and dark matter 
in the near future. In the theories where the neutrinos are Dirac particles, the Stueckelberg scenario and the theory based on $\U(1)_L$, we find a strong bound coming from the measurement of the number of relativistic species in the early Universe.  Our main results suggest that one could be optimistic about the possibility to test the different theories for neutrino masses if there is a simple connection to the properties and origin of dark matter candidates.

\section{NEUTRINO MASSES AND THE NEW PHYSICS SCALE}
\label{sec:NeutrinoMasses}
The observation of non-zero neutrino masses provides evidence that the Standard Model must be modified. At present, it remains unknown whether neutrinos are Dirac or Majorana fermions. 
In the scenario where neutrinos are Dirac in nature, their masses can be generated using the Yukawa interactions between the SM neutrinos, the SM Higgs and the additional right-handed neutrinos,
\begin{equation}
{\cal L}_\nu^D \supset Y_\nu \ \bar{\ell}_L i \sigma_2 H^* \nu_R + \text{h.c.}.
\end{equation}  
In order to generate neutrino masses in agreement with the experimental constraints, the Yukawa coupling $Y_\nu$ must be very small, i.e. if $Y_\nu \leq 10^{-12}$ then $m_\nu \leq 0.1$ eV. In this case it is necessary to forbid the Majorana mass term for the right-handed neutrinos which otherwise would be allowed by the SM gauge symmetries.

One of the simplest mechanisms to generate Majorana neutrino masses is the Type I seesaw mechanism \cite{Minkowski:1977sc,GellMann:1980vs,Yanagida:1979as,Mohapatra:1979ia}, where the following terms are added to the Lagrangian,
\begin{equation}
{\cal L}_\nu^M \supset Y_\nu \ \bar{\ell}_L  i \sigma_2 H^* \nu_R + \frac{1}{2} \nu^T_R C M_R \nu_R + \text{h.c.}, 
\end{equation}  
and once the right-handed neutrino masses are integrated out, the SM neutrino mass matrix is given by 
\begin{equation}
m_\nu = m_D M_R^{-1} m_D^T,
\end{equation}
where $m_D=Y_\nu v_0/\sqrt{2}$ and $v_0$ is the SM Higgs vacuum expectation value. If we take $m_D$ to be around the electroweak scale, $m_D \sim 10^2$ GeV, then we have that $M_R \lesssim 10^{14}$ GeV.  Therefore, in general, it will be difficult to test the theory for neutrino masses at current and future colliders.
However, we note that the masses $m_D$ and $M_R$ are unknown, and the seesaw mechanism could still be realized at a lower scale.

In simple gauge theories where the origin of neutrino masses can be understood, the seesaw scale is determined 
by the new gauge symmetry scale. The simplest theories are based on $\U(1)_{B-L}$ or $\U(1)_L$  \cite{FileviezPerez:2011pt,Duerr:2013dza,Perez:2014qfa,Schwaller:2013hqa}. In both scenarios, anomaly cancellation requires at least the addition of three right-handed neutrinos. Hence, the consistency of the theory automatically requires neutrino masses. The main difference between both possibilities is that, while in the case of $B-L$ the theory is already anomaly-free by the addition of three copies of right-handed neutrinos, in the case of $\U(1)_L$ one needs extra fields to define an anomaly-free theory. However, whereas in $\U(1)_{B-L}$ one is forced to include the dark matter by hand in such a way that the cancellation of the anomalies remains unspoiled, in the case of $\U(1)_L$ there is a natural dark matter candidate among the extra fermions that one needs to consider for anomaly cancellation.

\begin{figure}[t]
\centering
\includegraphics[width=.6\textwidth]{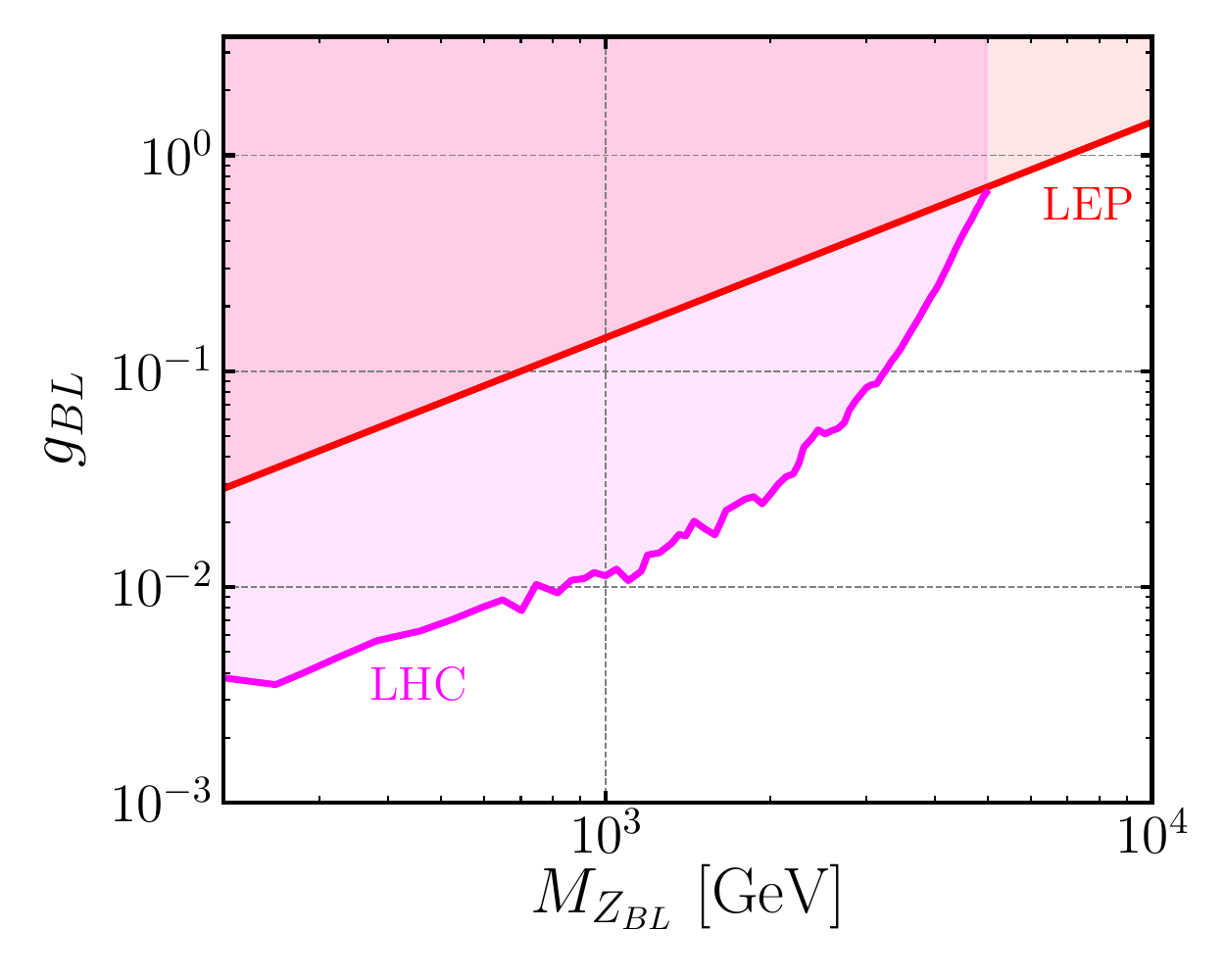} 
\caption{Summary of the collider bounds in the $\U(1)_{B-L}$ scenario in the $g_{BL}-M_{Z_{BL}}$ plane. The red line corresponds to the bound from LEP~\cite{Alioli:2017nzr}, while the pink line corresponds to dilepton searches at the LHC with $\sqrt{s}=13$ TeV and 36.1 fb$^{-1}$ \cite{Aaboud:2017buh}.
}
\label{fig:Collider_bounds}
\end{figure}

The LEP collider provides a bound on the $B\!-\!L$ gauge boson. In this work, we use the recent study of Ref.~\cite{Alioli:2017nzr} that gives
\beq
\frac{M_{Z_{BL}}}{g_{BL}} > 7 \,\, {\rm TeV}.
\eeq
This bound relies on the coupling with leptons, and hence, it can also be applied to the gauge boson of lepton number $Z_L$. Furthermore, dilepton searches at the LHC can also be used to constrain the $\text{U(1)}_{B-L}$ scenario, we use the result from ATLAS for center-of-mass energy $\sqrt{s}=13$ TeV and 36.1 fb$^{-1}$ of integrated luminosity \cite{Aaboud:2017buh}.  In Fig.~\ref{fig:Collider_bounds}, we show these bounds in the $g_{BL}-M_{Z_{BL}}$ plane. Here, $g_{BL}$ is the $B-L$ gauge coupling 
and $M_{Z_{BL}}$ is the mass of the new $B-L$ gauge boson. 
As can be appreciated, if the gauge coupling is of order 
one, $g_{BL} \approx 1$, the gauge boson must be heavier than 7-8 TeV. The bounds from the LHC become relevant in the region $M_{Z_{BL}} \lesssim 4$ TeV. 
In the next sections, we will discuss simple gauge theories where there is a link between the origin of neutrino masses and the properties 
of the DM candidate and show that the $B-L$ (or $L$) breaking scale must be, at most, in the multi-TeV scale.

\section{COSMOLOGICAL CONSTRAINTS: $N_{\rm eff}$ }
\label{sec:Neff}
New light states with non-negligible interactions with the SM could be copiously produced at high temperatures in the early Universe. During the radiation era, they would contribute to the total energy density of the Universe and therefore modify the predictions for the Cosmic Microwave Background (CMB). Therefore, the observation of the CMB with high angular resolution, as well as other indirect methods, such as the measurement of the abundance of light elements in the Universe, can impose relevant bounds on the existence of these light states. The measurement of the effective number of neutrino species can be used to constrain theories that have additional light particles that interact with the SM, see for example Refs.~\cite{Barger:2003zh,
Hamann:2011ge,
Anchordoqui:2011nh,
Hannestad:2012ky,
SolagurenBeascoa:2012cz,
Anchordoqui:2012qu,
Ho:2012ug,
Brust:2013xpv,
Boehm:2013jpa,
Perez:2013kla,
Escudero:2018mvt}, including right-handed neutrinos coupled to a $Z'$ and light thermal dark matter candidates. 
Axion-like particles that thermalize in the early Universe can also contribute to the value of $N_{\rm eff}$  \cite{Cadamuro:2010cz,
Salvio:2013iaa,
Kawasaki:2015ofa,
Baumann:2016wac}. 

\begin{figure}[t]
\centering
\includegraphics[width=0.445\linewidth]{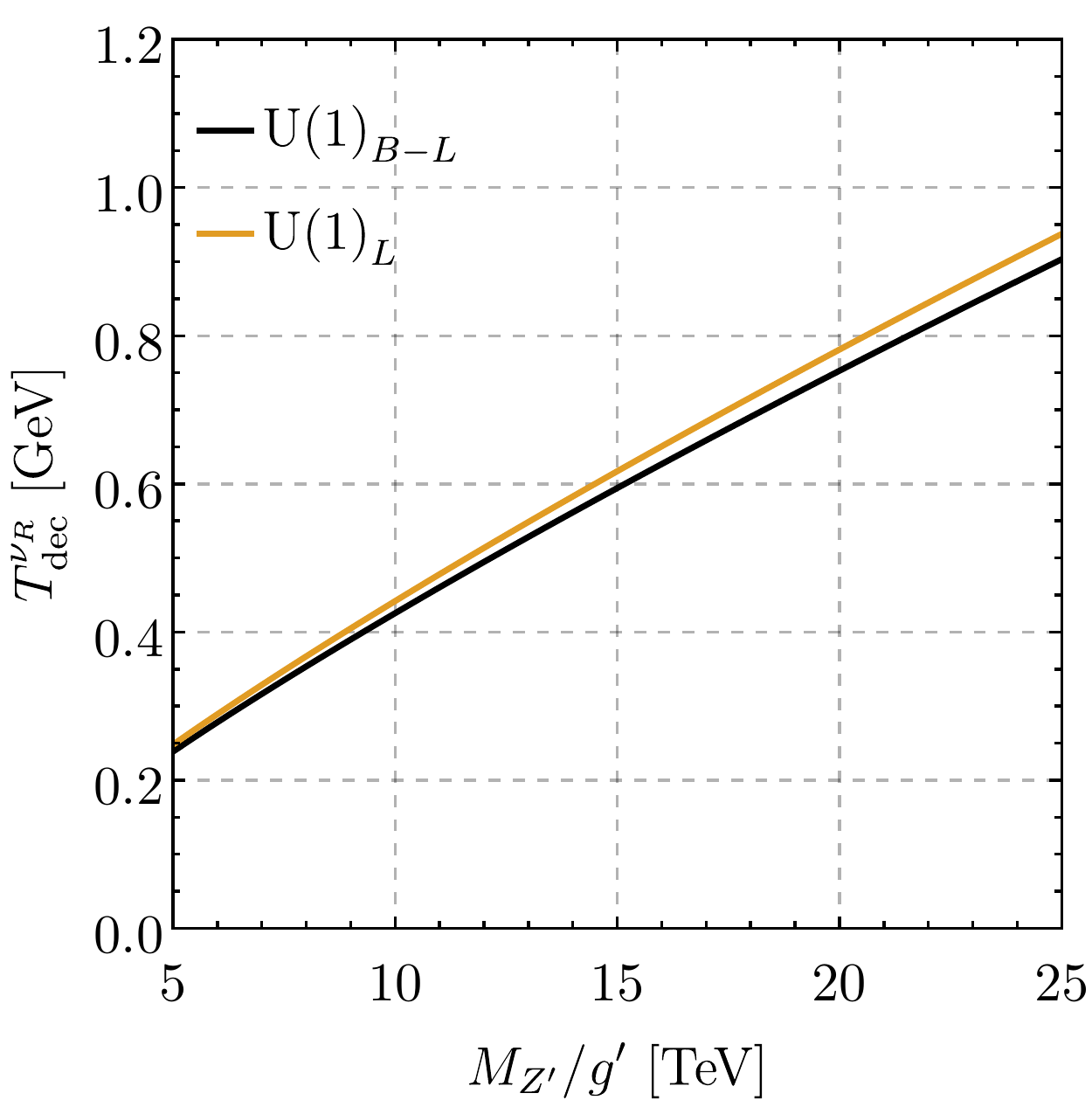} \,\,\,
\includegraphics[width=0.455\linewidth]{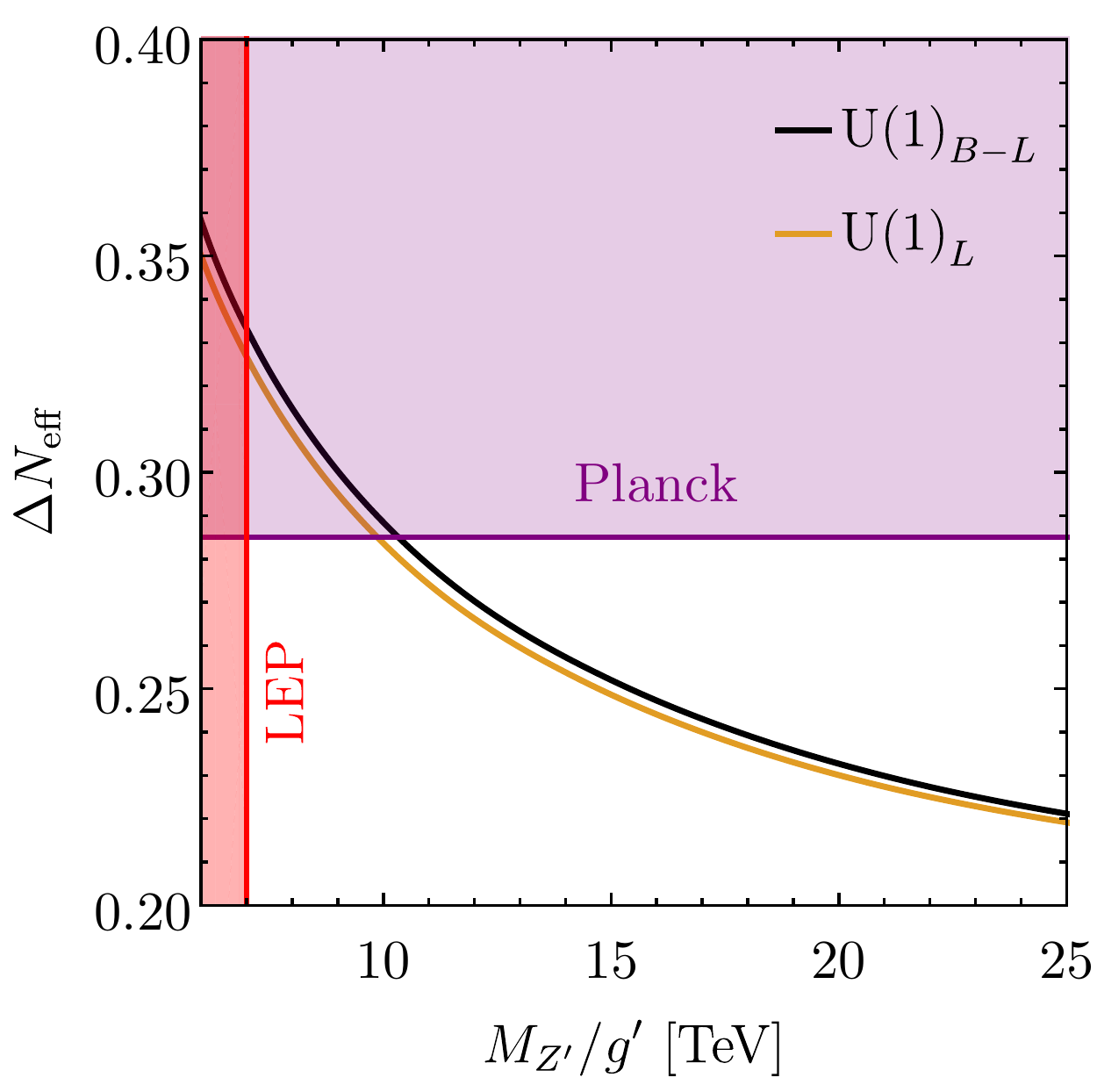} 
\caption{\textit{Left panel:} The decoupling temperature of the right-handed neutrinos $\nu_R$ as a function of $M_{Z'}/g'$. \textit{Right panel:} The effective number of extra relativistic species as a function of $M_{Z'}/g'$. The solid black and yellow lines correspond to the prediction for $\Delta N_{\rm eff}$ in the scenarios with $\U(1)_L$ and $\U(1)_{B-L}$, 
respectively. The region shaded in pink is excluded by the CMB measurement by the Planck satellite mission~\cite{Aghanim:2018eyx}. For comparison, we show the bound from LEP~\cite{Alioli:2017nzr} in red.}
\label{fig:Neff}
\end{figure}

In the scenarios in which the right-handed neutrinos $\nu_R$ are coupled to a new gauge boson they could thermalize and contribute to the effective number of neutrino species as
\begin{equation}
\Delta N_\text{eff} = N_\text{eff} - N_\text{eff}^\text{SM} = N_{\nu_R} \left(\frac{T_{\nu_R}}{T_{\nu_L}}\right)^4 = N_{\nu_R} \left(\frac{g(T_{\nu_L}^\text{dec})}{g(T_{\nu_R}^\text{dec})}\right)^\frac{4}{3},
\end{equation}
where the last equality follows from conservation of entropy in the plasma and $g(T)$ corresponds to the relativistic degrees of freedom at temperature $T$. Here, $N_{\nu_R}$ refers to the number of relativistic right-handed neutrinos. For the active neutrinos we have $T_{\nu_L}^\text{dec}\approx 2.3$ MeV \cite{Enqvist:1991gx} and hence $g(T_{\nu_L}^\text{dec})=43/4$. For the prediction in the SM we take the recent result $N_\text{eff}^\text{SM} = 3.045$ \cite{deSalas:2016ztq}.

In order to predict the shift in the effective number of neutrino species, $\Delta N_\text{eff}$, one needs to estimate the temperature at which the right-handed neutrinos decouple from the plasma. The latter occurs when the interaction rate drops below the expansion rate of the Universe,
\begin{equation}
\Gamma(T_{\nu_R}^\text{dec}) = H(T_{\nu_R}^\text{dec}),
\end{equation}
where the Hubble expansion parameter is
\begin{equation}
H(T) = \sqrt{\frac{8 \pi G_N \rho(T)}{3}}=\sqrt{\frac{4\pi^3 G_N}{45} \left( g(T) +   \displaystyle 3\frac{7}{8} g_{\nu_R} \right)}\, \,T^2,
\end{equation}
where $g_{\nu_R}=2$ is the number of spin states of the right-handed neutrinos and $g(T)$ is the number of relativistic SM species in thermal equilibrium at temperature $T$. In the calculation of $g(T)$, one needs to take into account the QCD phase transition, i.e. the threshold between quarks and hadrons as degrees of freedom. This transition can be computed via lattice QCD. Here we will use the results from Ref.~\cite{Borsanyi:2016ksw}. 

The right-handed neutrinos remain in thermal equilibrium with the SM via exchange of a new gauge boson $Z'$,
\begin{eqnarray}
\Gamma_{\nu_R} (T) &=& n_{\nu_R}(T) \langle \sigma (\bar{\nu}_R \nu_R \to \bar{f} f ) \, v_M \rangle \\
&=&\frac{g_{\nu_R}^2}{n_{\nu_R}(T)}\int \frac{d^3\vec{p}}{(2\pi)^3} f_{\nu_R}(p) \int \frac{d^3\vec{k}}{(2\pi)^3} f_{\nu_R}(k) \sigma_f(s) v_M, \nonumber 
\end{eqnarray}
where the Fermi-Dirac distribution has been used to determine the number density of particles in a fermion gas at the thermal equilibrium,
\begin{equation}
f_{\nu_R} (k)=(e^{(k-\mu)/T}+1)^{-1},
\end{equation}
and since the particles in the gas are relativistic, the chemical potential $\mu$ can be ignored. In the above equation, $v_M = (1-\cos \theta)$ refers to the Moller velocity and $s = 2 p k (1-\cos \theta)$ where $p$ and $k$ are the momenta of the interacting relativistic particles and $\theta$ is the angle between them.
For a massless right-handed neutrino,
\begin{equation}
n_{\nu_R}(T) = g_{\nu_R} \int \frac{d^3 \vec{k}}{(2\pi)^3}f_\nu(k) = \frac{3}{2\pi^2}\xi(3) T^3.
\end{equation}
The cross-section for this process is given by 
\begin{equation}
\sigma_{ \bar{\nu}_R\nu_R\to \bar{f}f } = \frac{g'^4}{12 \pi \sqrt{s}}\frac{1}{(s-M_{Z'}^2)^2 + \Gamma_{Z'}^2M_{Z'}^2}\sum_f N_f^C  n_f^2\sqrt{s-4M_f^2}\, (2M_f^2+s),
\end{equation}
where $N_f^C$ is the color multiplicity of the fermion, i.e. $N_f^C = 1 (3)$ for leptons (quarks) and $n_f$ is the charge of the fermion under the new symmetry through which the right-handed neutrinos interact with the rest of the plasma. In this work, we focus on heavy mediators $T^{\rm dec}_{\nu_R} \ll M_{Z'}$, and hence, we can work in the limit $s \ll M_{Z'}$. Neglecting the fermion masses, the interaction rate reads
\begin{equation}
\Gamma_{\nu_R} (T) = \frac{49 \pi^5 T^5}{97200 \xi(3)}\left(\frac{g'}{M_{Z'}}\right)^4 \sum_f N_{f}^C n_f^2,
\end{equation}
where the sum is performed over all SM fermions that are in thermal equilibrium at the temperature $T$. In order to understand the bounds in models where the right-handed neutrinos are very light, we use the value for $N_{\rm eff}$ derived from the CMB measurement by the Planck satellite mission \cite{Aghanim:2018eyx},
\begin{align}
N_\text{eff} &= 2.99^{+0.34}_{-0.33} \hspace{0.3cm} \Rightarrow \hspace{0.3cm} \Delta N_\text{eff} < 0.285,
\end{align}
adopting the most conservative limit. Moreover, future CMB Stage-IV experiments \cite{Abazajian:2016yjj} are expected to improve this measurement to $\Delta N_\text{eff} < 0.06$.

In theories based on local $B\!-\!L$ or $L$ gauge symmetries that have Dirac neutrinos, the right-handed counterparts of the latter thermalize with the SM and contribute to $N_{\rm eff}$. 
In the left panel of Fig.~\ref{fig:Neff} we present our results for the decoupling temperature of the right-handed neutrinos, $\nu_R$, as a function of the ratio $M_{Z'}/g'$. 
In the right panel of Fig.~\ref{fig:Neff} we present the prediction for $\Delta N_{\rm eff}$ in these models. As can be seen in the figure, in these theories the bound coming from Planck is stronger than the collider bound coming from LEP. Basically, in these theories we find that $M_{Z^{'}}/g^{'} > (9-10)$ TeV in order to satisfy the $N_{\rm eff}$ bounds. This has implications for any gauge theory coupled to the SM with very light new particles. In the next sections, we will study the implications of this bound on the phenomenology of dark matter candidates.

\section{$B-L$ GAUGE THEORIES, NEUTRINOS AND DARK MATTER}
\label{sec:B-L}

\subsection{Stueckelberg Scenario}
\label{sec:Stueckelberg}
Neutrinos are Dirac particles if $B\!-\!L$ is a conserved symmetry. The local gauge symmetry $\U(1)_{B-L}$ can remain conserved, and the associated gauge boson can acquire mass through the Stueckelberg mechanism. 
For a review on the Stueckelberg mechanism, see Refs.~\cite{Ruegg:2003ps,Feldman:2011ms}. Due to the absence of a scalar particle in the $B\!-\!L$ sector, it is not possible to write a mass term for a Majorana fermion with $B\!-\!L$ charge. Therefore, the DM could be a scalar $\phi$ or a Dirac fermion  with quantum numbers $\chi \sim (1,1,0,n_\chi)$, corresponding to the gauge groups ($\SU(3)_c$, $\SU(2)_L$, $\U(1)_Y$, $\U(1)_{B-L}$). The state $\chi$ is stable for $n_\chi \neq 1$; since this value allows for $\chi$ to mix with the neutrinos and decay. In addition, $n_\chi$ cannot be arbitrarily large; perturbativity requires $n_\chi\cdot g_{BL}<\sqrt{4\pi}$ to be satisfied.
In the case where DM is a scalar field, it is not possible to predict an upper bound for the mass of the $B\!-\!L$ gauge boson. This is because, through the introduction of a Higgs portal term, $\phi^\dagger \phi H^\dagger H$, the DM can be produced and the $Z_{BL}$ boson can be decoupled. Thus, we discuss the scenario where the DM is a Dirac fermion. In this case, the relevant Lagrangian is given by
 \begin{align}
\mathcal{L}_D \supset & \  i \overline{\chi}_L \gamma^\mu D_\mu \chi_L + i \overline{\chi}_R \gamma^\mu D_\mu \chi_R 
- (Y_\nu \overline{l}_L i \sigma_2 H^* \nu_R + M_\chi \overline{\chi}_L \chi_R + {\rm h.c.})  \nonumber \\
& - \frac{1}{2} \left( M_{Z_{BL}} Z^\mu_{BL} + \partial^\mu \sigma \right)  \left( M_{Z_{BL}} Z_{\mu BL} + \partial_\mu \sigma \right), 
\end{align}   
%
which is invariant under the following gauge transformations, 
\beq
\delta Z_{BL}^\mu=\partial^\mu \lambda(x) \hspace{5mm} {\rm and} \hspace{5mm} \delta \sigma = -M_{Z_{BL}} \lambda(x),
\eeq
and the $\sigma$ field decouples from the theory.
When the Stueckelberg mechanism is applied to an Abelian gauge group, the theory is renormalizable and unitary.  However, for non-Abelian gauge groups, violation of unitarity arises at tree-level in the scattering of longitudinal gauge bosons.

\begin{figure}[tbp]
\centering
\includegraphics[width=.495\textwidth]{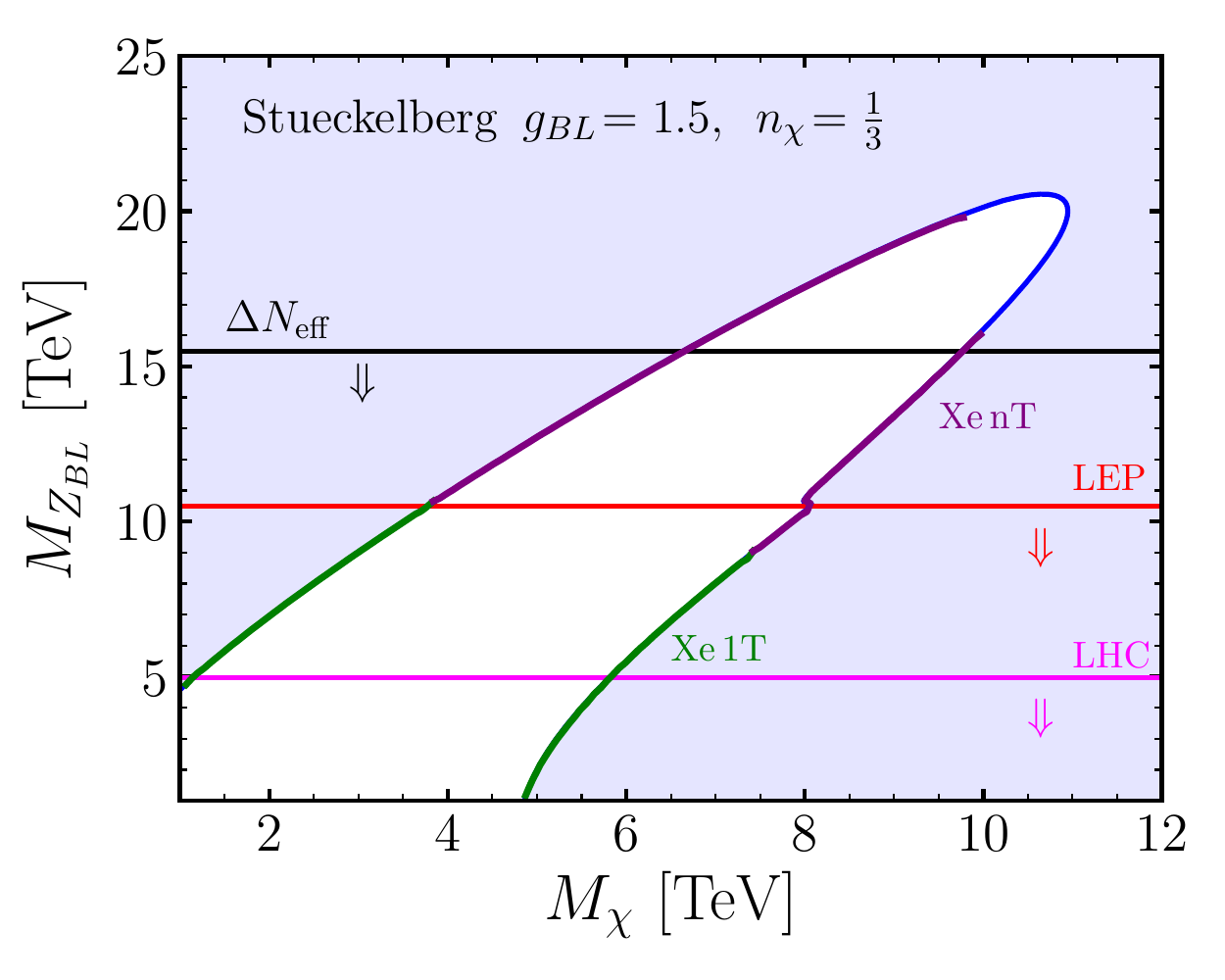} 
\includegraphics[width=.495\textwidth]{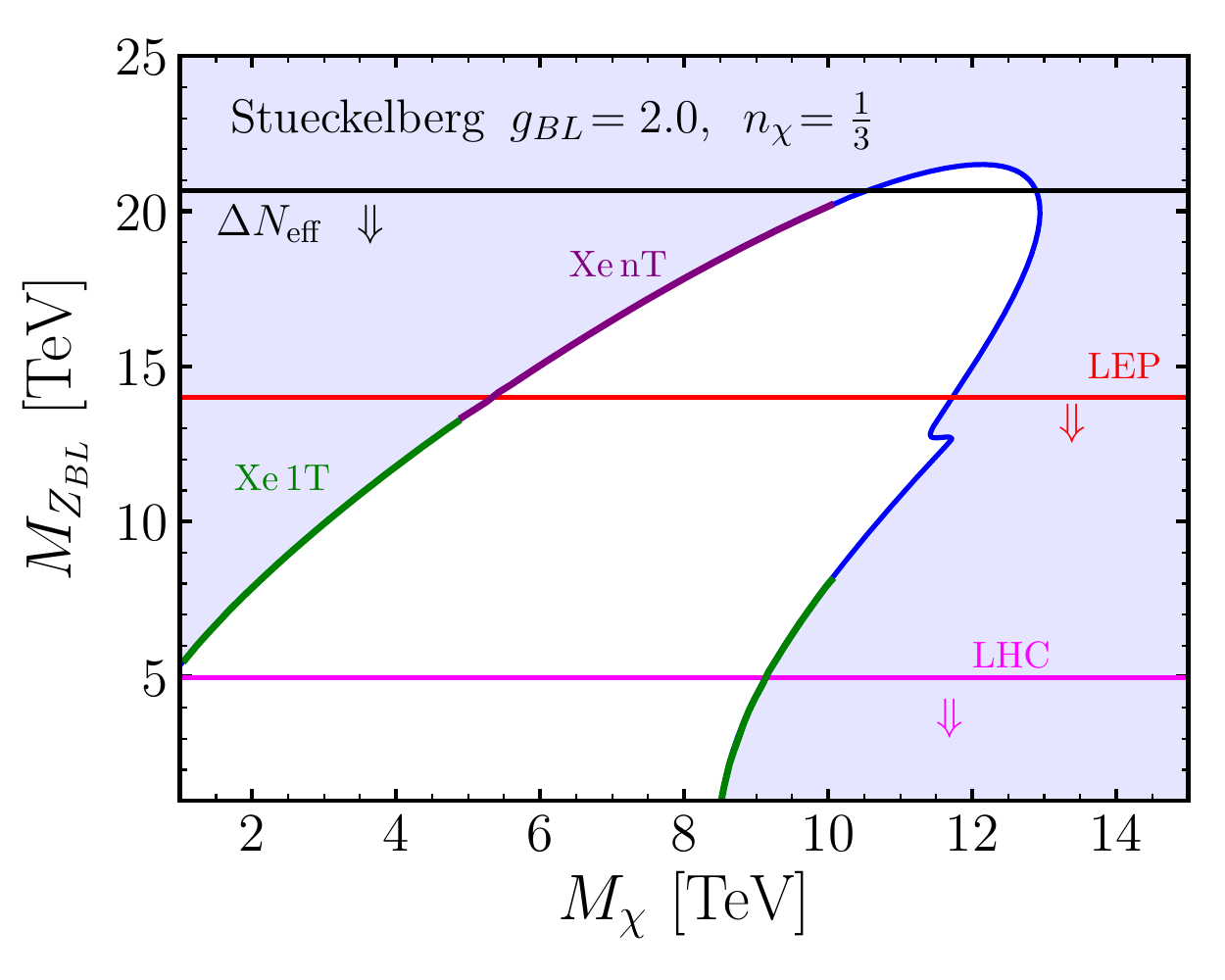} 
\caption{
Results for the dark matter relic density in the $(M_\chi,\, M_{Z_{BL}})$ plane for the Stueckelberg scenario. The solid blue line satisfies the correct dark matter relic abundance $\Omega_\chi h^2 =0.1200 \pm 0.0012$~\cite{Aghanim:2018eyx}, while the region shaded in light blue overproduces it. 
For the points that saturate the relic abundance, we mark in green those that are excluded by Xenon-1T \cite{Aprile:2017iyp} and in purple those that will be reached by the Xenon-nT \cite{Aprile:2015uzo} experiment. 
The horizontal red (pink) line corresponds to the LEP~\cite{Alioli:2017nzr}  (LHC~\cite{Aaboud:2017buh}) bound on the mass of the $Z_{BL}$ gauge boson. The left (right) panel corresponds to  $g_{BL}\!=\! 1.5$ ($g_{BL}\!=2$ ) and $n_\chi=1/3$.  The bounds from $N_{\rm eff}$ are shown by the horizontal black line.}
\label{fig:mdmStueckelberg}
\end{figure}

The properties of the dark matter candidate, $\chi=\chi_L + \chi_R$, are defined 
by the $B-L$ gauge interaction. The model contains only four free parameters
\beq
M_\chi, \,\,\,\, M_{Z_{BL}}, \,\,\,\, g_{BL}, \,\,\,\, {\rm{and}} \ n_\chi.
\eeq
For the rest of our discussion we fix $n_\chi\!=\!1/3$ for simplicity, but the main conclusions can be applied to any other scenario with different charge. The annihilation channels of our dark matter candidate are:
$$\bar{\chi} \chi \to e_i^+ e_i^-, \, \bar{\nu}_i \nu_i, \, \bar{u}_i u_i, \, \bar{d}_i d_i, \, Z_{BL} Z_{BL},$$
see Appendix \ref{sec:AppendixB} for their explicit representations in Feynman graphs. We note that the annihilation channel into fermions is the dominant one, with a higher contribution from annihilation into leptons due to their larger coupling to $Z_{BL}$;  the leptons have $B\!-\!L$ charge $-1$, while quarks have $+1/3$. 
In this model, the perturbative bound on the gauge coupling is given by $g_{BL} < 2 \sqrt{\pi}$.

In order to compute numerically the dark matter relic abundance, $\Omega_\chi h^2$,  we use \texttt{MicrOMEGAs 5.0.6} \cite{Belanger:2018mqt}, implementing the model with the help of \texttt{LanHEP 3.2} \cite{Semenov:2014rea}. We cross-check our results with an independent calculation in \texttt{Mathematica}.
In Fig.~\ref{fig:mdmStueckelberg}, we present our results in the $(M_\chi,\, M_{Z_{BL}})$ plane for the Stueckelberg scenario. The solid blue line satisfies the correct dark matter relic abundance $\Omega_\chi h^2 =0.1200 \pm 0.0012$~\cite{Aghanim:2018eyx}, while the region shaded in light blue overproduces it. The latter is then ruled out by cosmology unless the thermal history of the Universe is altered. The horizontal red (pink) line corresponds to the LEP~\cite{Alioli:2017nzr} (LHC~\cite{Aaboud:2017buh}) bound on the mass of the $Z_{BL}$ gauge boson. The left (right) panel corresponds to  $g_{BL}\!=\! 1.5$ ($g_{BL}\!=2$ ) and $n_\chi=1/3$. The solid green line shows current experimental bounds from Xenon-1T \cite{Aprile:2017iyp} and the purple lines shows the projected sensitivity for Xenon-nT \cite{Aprile:2015uzo}. The small feature that can be observed in the right panel at around $M_\chi \approx 12$ TeV corresponds to the region in parameter space where the $\chi \chi \rightarrow Z_{BL} Z_{BL}$  channel also contributes to the relic density.

The right-handed neutrinos feel the $B\!-\!L$ interaction and they could be thermalized with the SM plasma in the early Universe and contribute to the effective number of neutrino species. Therefore, the bound on $\Delta N_{\rm eff}$, discussed in Section~\ref{sec:Neff}, should be taken into account. This bound corresponds to
\beq
\Delta N_{\rm eff} < 0.285 \hspace{0.5cm} \Rightarrow \hspace{0.5cm} \frac{M_{Z_{BL}}}{g_{BL}} > 10.33 \,\, {\rm TeV},
\eeq
and is given by the black line in Fig.~\ref{fig:mdmStueckelberg}. We note that, as the figure shows, this bound is stronger than the LEP bound. As can be appreciated, the upper bound on the gauge coupling for $n_\chi = 1/3$ is $g_{BL} \lesssim 2$, since scenarios with larger values for the gauge coupling are totally excluded by the bounds on $N_\text{eff}$. Therefore, using all cosmological bounds one finds an upper bound on the gauge boson mass, i.e. $M_{Z_{BL}} \lesssim 22$ TeV. 

Direct detection experiments aim to measure the nuclear recoil from interaction with a dark matter particle. In this model, the only interaction between dark matter and the nucleon in atoms occurs via the exchange of a gauge boson,
$$N \chi \to Z_{BL}^* \to N \chi.$$ 
The direct detection spin-independent cross-section is then given by,
\beq
\label{eq:sigmaSI}
\sigma_{\rm SI} = \frac{m_N^2  M_{\rm \chi}^2}{\pi (m_N + M_{\rm \chi})^2} \frac{n_\chi^2 g_{BL}^4}{M_{Z_{BL}}^4} , 
\eeq
where $m_N$ corresponds to the nucleon mass.

\begin{figure}[t]
\centering
\includegraphics[width=.7\textwidth]{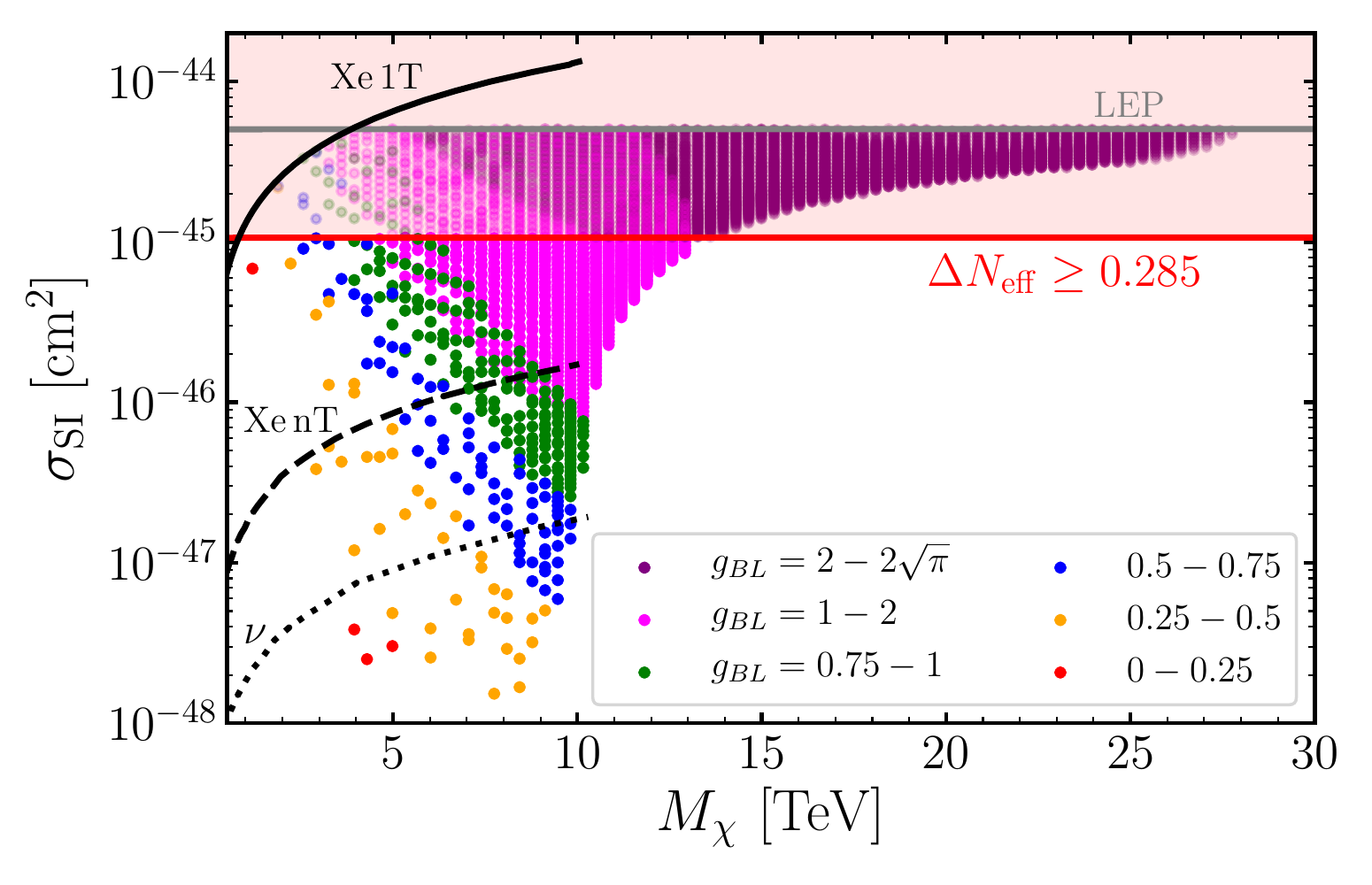} 
\caption{Predictions for the direct detection spin-independent cross-section as a function of the dark matter mass in the Stueckelberg $\U(1)_{B-L}$ scenario. 
Points with different colors correspond to different values of the gauge coupling as shown in the legend.
All points shown here satisfy the correct relic density $\Omega_\chi h^2 =0.1200 \pm 0.0036$. The solid black line shows current experimental bounds from Xenon-1T~\cite{Aprile:2017iyp}, the dashed black line shows the projected sensitivity for Xenon-nT \cite{Aprile:2015uzo}, and the dotted black line shows the coherent neutrino scattering limit~\cite{Billard:2013qya}. The bounds from $N_{\rm eff}$ are shown by a red horizontal line, ruling out the parameter space above this line.
}
\label{fig:DDStueckelberg}
\end{figure}

In Fig.~\ref{fig:DDStueckelberg}, we show the predictions for the direct detection spin-independent cross-section as a function of the dark matter mass in the Stueckelberg $\U(1)_{B-L}$ scenario. Red points correspond to $g_{BL}=0-0.25$, orange points correspond to $g_{BL}=0.25-0.5$, blue points correspond to $g_{BL}=0.5-0.75$, green points correspond to $g_{BL}=0.75-1.0$, pink points correspond to $g_{BL}=1.0-2.0$ and purple points correspond to $g_{BL}=2.0-2\sqrt{\pi}$. The solid black line shows current experimental bounds from Xenon-1T \cite{Aprile:2017iyp}, the dashed black line shows the projected sensitivity for Xenon-nT \cite{Aprile:2015uzo} and the dotted black line shows the coherent neutrino scattering limit~\cite{Billard:2013qya}. In this figure, we also present the bound coming from $\Delta N_{\rm eff}$. As it can be seen, this is a strong bound, and it excludes a large region in parameter space that otherwise could be reached by Xenon-nT. However, some of the points lie below the neutrino floor and hence will escape detection from future direct detection experiments. 

In the scenario we have considered,
the projected sensitivity of CMB Stage-IV for the measurement of $N_{\rm eff}$ will fully probe this model. However, it should be noted that the bound coming from $\Delta N_{\rm eff}$ can be relaxed by taking the limit $n_\chi \gg 1$, since this would require $g_{BL}\ll 1$ for the dark matter relic density to be explained. Even though the bounds on $N_\text{eff}$ may change under the choice of a different $n_\chi$, it can be shown that the upper bound on the $B\!-\!L$ scale will not go beyond $100$ TeV regardless of the choice of the charge~\cite{FileviezPerez:2018toq}. For the case considered here, one can also find an upper bound on the dark matter mass, i.e. $M_{\chi} \lesssim 13$ TeV. Then, there is hope to test or rule out this simple theory for neutrino masses and dark matter in the near future.

\subsection{Canonical Seesaw Scenario}
Majorana neutrino masses can be generated through the spontaneous breaking of the $B-L$ symmetry in two units, i.e. we introduce a new Higgs with quantum numbers $S_{BL} \sim (1,1,0,2)$. In this section, we will investigate the different dark matter scenarios in the case where the neutrinos are Majorana particles and their masses are generated through the canonical Type I seesaw scenario. In contrast to the Stueckelberg scenario, these models predict violation of lepton number. 

The DM can be a Dirac fermion if we add a pair of vector-like fermionic fields, i.e. $\chi_L \sim (1,1,0,n_\chi)$ and $\chi_R \sim (1,1,0,n_\chi)$, where $n_\chi \neq 1,3$; in order to avoid mixing with neutrinos and avoid the decay of DM. If we allow for non-renormalizable operators, odd values of $n_\chi$ will give mixing between the dark matter candidate and neutrinos, and hence should be forbidden. 
The Lagrangian in this case is given by,
\begin{align}
\mathcal{L} \supset &  \ i \overline{\chi}_L \gamma^\mu D_\mu \chi_L + i \overline{\chi}_R \gamma^\mu D_\mu \chi_R + (D_\mu S_{BL})^\dagger (D^\mu S_{BL}) \nonumber \\ 
& - (Y_\nu \overline{l}_L i \sigma_2 H^* \nu_R + y_R \, \nu_R^T C \nu_R S_{BL} + M_\chi \overline{\chi}_L \chi_R + {\rm h.c.}) . 
\end{align}
The covariant derivative for $\chi$ is given by $D^\mu \chi_L=\partial^\mu \chi_{L} + ig_{BL} n_\chi Z_{BL}^\mu \chi_{L}$, and similarly for $\chi_R$.
The scalar potential is given by,
\begin{align}
V(H, S_{BL}) = & -\mu_H^2 H^\dagger H - \mu_{BL}^2 S_{BL}^\dagger S_{BL} + \lambda_H (H^\dagger H )^ 2  + \lambda_{BL} (S_{BL}^\dagger S_{BL} )^2 + \lambda_{HBL} (H^\dagger H) (S_{BL}^\dagger S_{BL}), 
\end{align}
where $H$ corresponds to the SM Higgs doublet and $S_{BL}$ is the new Higgs, only charged under the $\U(1)_{B\!-\!L}$ group. In the zero temperature vacuum of the theory, both fields acquire a non-zero vacuum expectation value and we can write, 
\beq
S_{BL}=\frac{1}{\sqrt{2}} \left( s_{BL}+v_{BL} \right), \quad \text{ and } \quad  H = \frac{1}{\sqrt{2}}\begin{pmatrix}
  0 \\  h+v_H
 \end{pmatrix},
\eeq
where the Higgs doublet has been written in the unitary gauge. This leads to mixing among both scalars, and hence, the mass matrix needs to be diagonalized in order to find the physical states. The latter are given by,
\begin{align}
h_1 &= h \cos \theta_{BL} - s_{BL} \sin \theta_{BL}, \\
h_2 &= s_{BL} \cos\theta_{BL} + h \sin \theta_{BL},
\end{align}
where the scalar mixing angle can be written in terms of the scalar quartic couplings and the vevs,
\beq
\tan 2\theta_{BL}  = \frac{\lambda_{HBL} \, v_H v_{BL}} {\lambda_{BL} v_{BL}^2  - \lambda_H v_H^2}.
\eeq

After spontaneous symmetry breaking, the $B\!-\!L$ gauge boson and the right-handed neutrinos acquire the following masses,
\beq
M_{Z_{BL}} = 2 g_{BL} v_{BL},  \hspace{6mm} {\rm{and}} \hspace{6mm} M_R = \sqrt{2} \, y_R v_{BL}.
\eeq
In this theory, the Majorana neutrino masses are generated through the Type I seesaw mechanism 
and the right-handed neutrinos are around the TeV scale unless very small Yukawa couplings, $y_R$, are assumed.
In this model, the perturbative bound on the gauge coupling comes from the 
$S_{BL}^\dagger S_{BL} Z_{BL} Z_{BL}$ interaction when $n_\chi \leq 2$ and it is therefore given by $g_{BL} \leq \sqrt{\frac{\pi}{2}}$.

Henceforth, we set the SM Higgs boson mass to $M_{h_1}=125.09$ GeV and $v_H=246.22$ GeV. We also set the masses of the three right-handed neutrinos to the same value $M_R$ without loss of generality.
Then, the model contains seven free parameters
\beq
M_\chi, \,\,\,\, M_{Z_{BL}}, \,\,\,\, M_R, \,\,\,\, M_{h_2}, \,\,\,\, \theta_{BL}, \,\,\,\, g_{BL}, \  {\rm{and}} \ n_\chi.
\eeq
For the rest of this section we fix the dark matter $B\!-\!L$ charge to $n_\chi\!=\!1/3$.
The rest of the parameters in the Lagrangian can be expressed as a function of them,
\begin{align}
\mu_H^2 = & \lambda_H v_H^2 + \frac{\lambda_{HBL}}{2} v_{BL}^2, \\[1ex]
\mu_{BL}^2 = & \lambda_{BL} v_{BL}^2 + \frac{\lambda_{HBL}}{2} v_H^2, \\
\lambda_H = & \frac{1}{2v_H^2} \left( M_{h_1}^2 \cos^2 \theta_{BL} + M_{h_2}^2 \sin^2 \theta_{BL} \right),\\
\lambda_{BL} = & \frac{1}{2v_{BL}^2} \left( M_{h_1}^2 \sin^2 \theta_{BL} + M_{h_2}^2 \cos^2 \theta_{BL} \right),\\
\lambda_{HBL} = & \frac{1}{v_H v_{BL}} \left( M_{h_2}^2  - M_{h_1}^2 \right) \sin \theta_{BL} \cos \theta_{BL}  .
\end{align}
To ensure vacuum stability of the scalar potential we impose
\begin{align}
\lambda_H, \lambda_{BL} > 0 \hspace{5mm} {\rm and} \hspace{5mm} \lambda_{HBL} > -2\sqrt{\lambda_H \lambda_{BL}},
\end{align}
and we also check for the perturbativity of the couplings $\lambda_i,g_{BL} < \sqrt{4\pi}$. 
\begin{figure}[t]
\centering
\includegraphics[width=.55\textwidth]{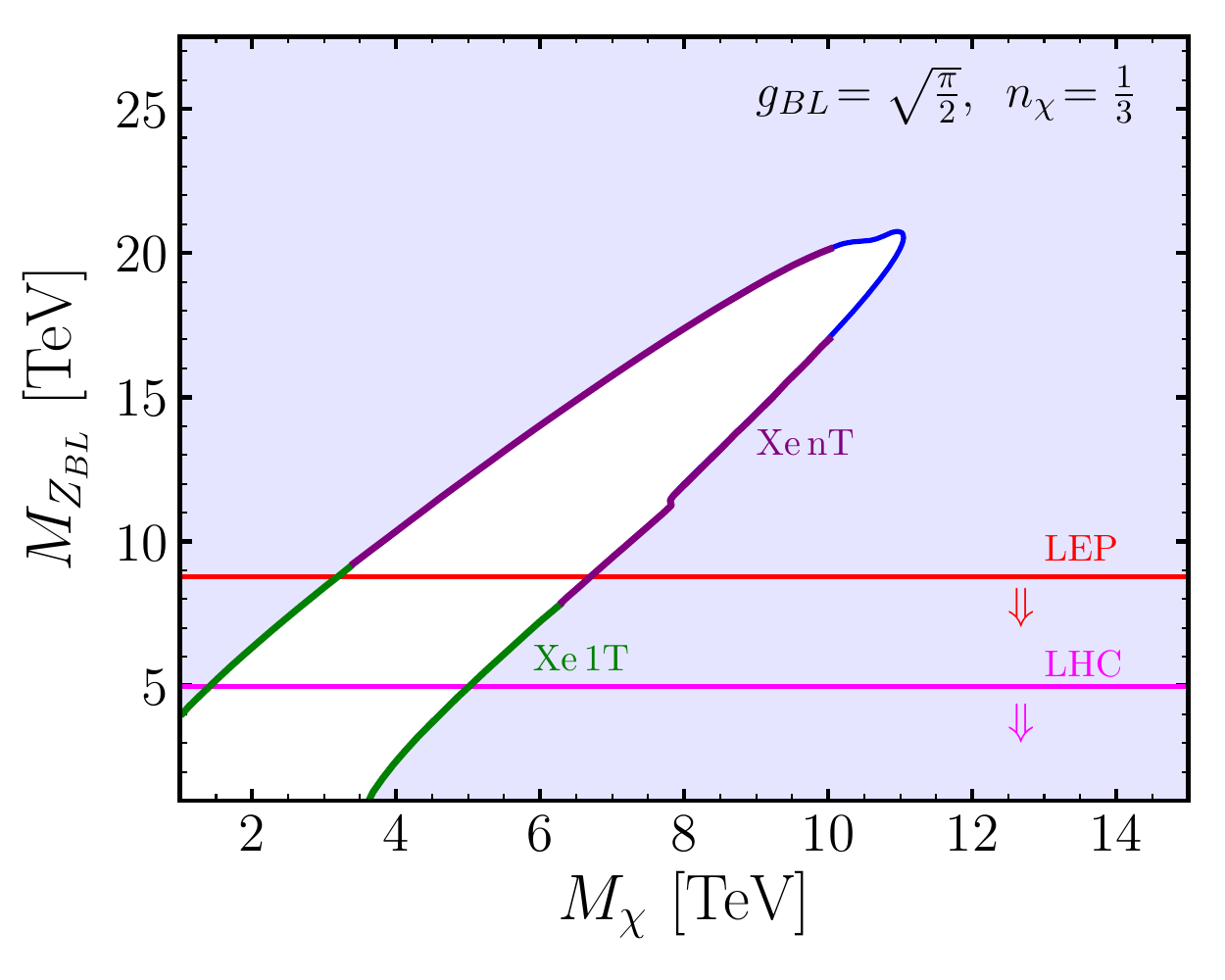} 
\caption{Results for the dark matter relic density in the $(M_{\chi},\,M_{Z_{BL}})$ plane. The solid blue line satisfies the correct dark matter relic abundance $\Omega_\chi h^2 = 0.1200 \pm 0.0012$~\cite{Aghanim:2018eyx}, while the region shaded in light blue overproduces it. For the points that saturate the relic abundance, we mark in green those that are excluded by Xenon-1T \cite{Aprile:2017iyp} and in purple those that will be reached by the Xenon-nT \cite{Aprile:2015uzo} experiment. The horizontal red (pink) line corresponds to the LEP~\cite{Alioli:2017nzr}  (LHC~\cite{Aaboud:2017buh}) bound on the mass of the $Z_{BL}$ gauge boson. We fix $n_\chi=1/3$, $M_R\!=\!M_{h_2}=1$ TeV and $g_{BL}\!=\!\sqrt{\pi/2}$. 
Here we assume no mixing between the two Higgses in the theory, $\theta_{BL}=0$.}
\label{bounds}
\end{figure}
The annihilation channels of our DM candidate in this theory are
$$\bar{\chi} \chi \to e_i^+ e_i^-, \, \bar{\nu}_i \nu_i, \, \bar{u}_i u_i, \, \bar{d}_i d_i, \, \bar{N}_i N_i, \, Z_{BL} Z_{BL}, \ Z_{BL} h_i,$$ 
whose explicit Feynman graphs are shown in Appendix~\ref{sec:AppendixB}. Here, $h_i=h_1,h_2$ are the Higgses present in the theory.

In Fig.~\ref{bounds}, we show the allowed parameter space by dark matter relic density in the $(M_{\chi},\,M_{Z_{BL}})$ plane for the maximal value of the gauge coupling. The solid blue line satisfies the correct dark matter relic abundance $\Omega_\chi h^2 = 0.1200 \pm 0.0012$, while the region shaded in light blue overproduces it. The horizontal red (pink) line corresponds to the LEP (LHC) bound on the mass of the $Z_{BL}$ gauge boson. We set $M_R\!=\!M_{h_2}=1$ TeV, $g_{BL}\!=\!\sqrt{\pi/2}$ and zero scalar mixing angle. In this model, the dark matter candidate has no Yukawa interaction with $S_{BL}$, and hence, there is no Higgs portal between the DM and the SM fermions. However, a small value for $\theta_{BL}$ will only have a small impact on the calculation of the dark matter relic density. For the values of $M_{h_2}$ we consider, we take the bound $\sin \theta_{BL}\leq0.3$ \cite{Ilnicka:2018def}.

\begin{figure}[h]
\centering
\includegraphics[width=.495\textwidth]{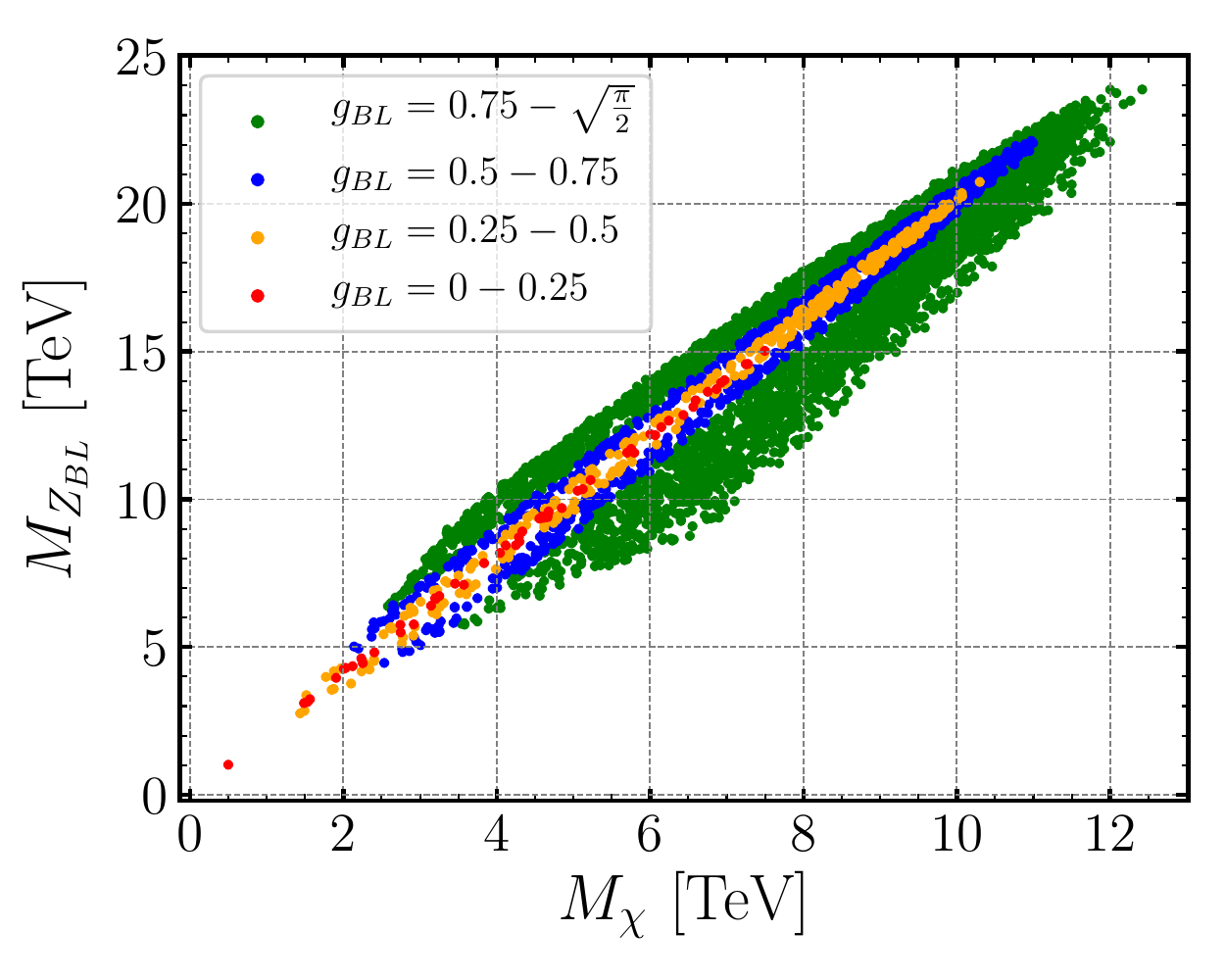} 
\includegraphics[width=.495\textwidth]{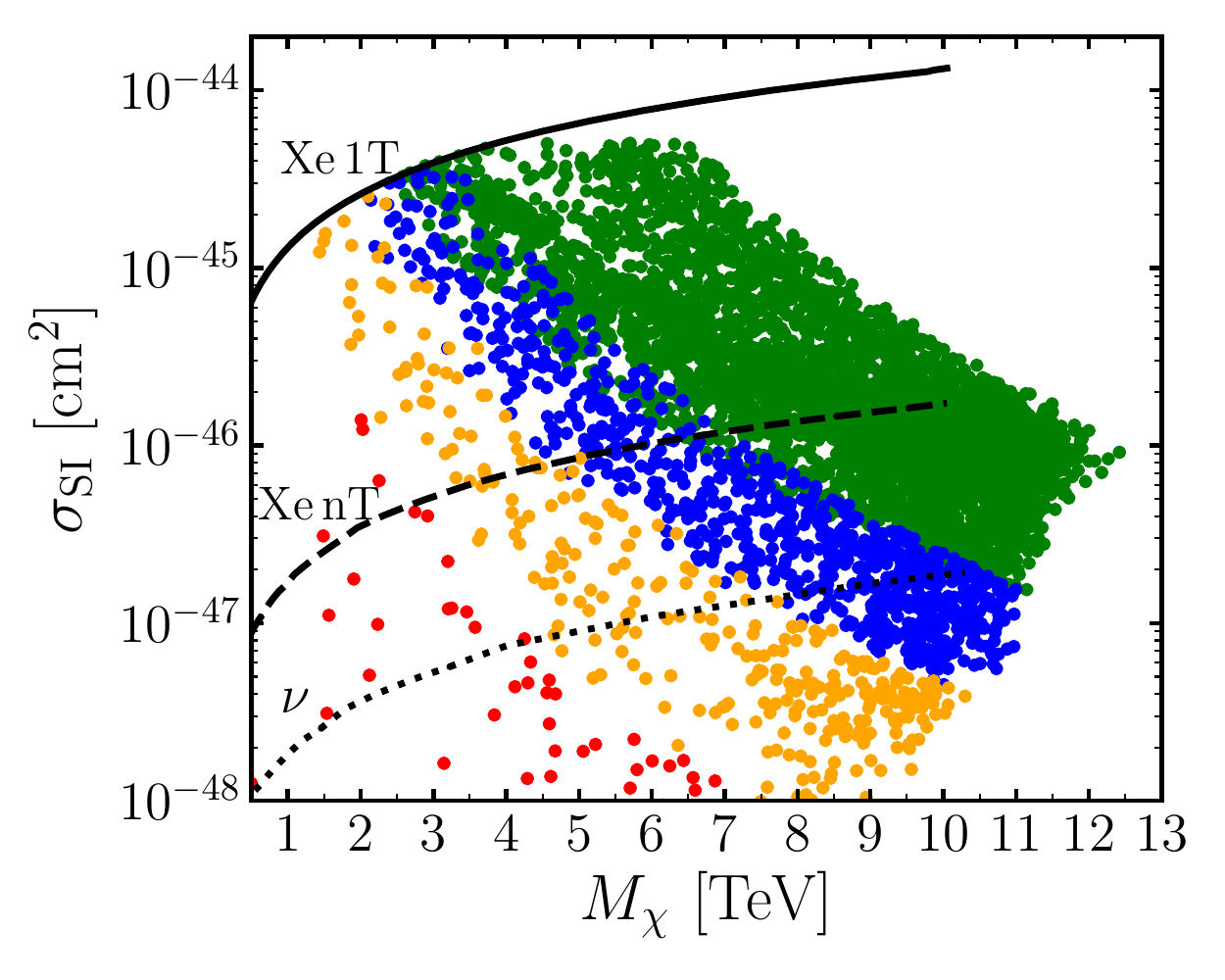} 
\caption{
\textit{Left panel:} Scatter plot of points with $n_\chi\!=\!1/3$ in the $M_{Z_{BL}}-M_\chi$ plane that are in agreement with the measured relic abundance, $\Omega_\chi h^2 = 0.1200 \pm 0.0036$. Points with different colors correspond to different values of the gauge coupling as shown in the legend. All points shown satisfy bounds from direct detection and LEP~\cite{Alioli:2017nzr}. \textit{Right panel:} Predictions for the direct detection spin-independent cross-section as a function of the dark matter mass, for the same points as in the left panel. We perform a random scan on $M_{h_2}$ and $M_R$ in the range $[0.1-20]$ TeV. For the scalar mixing angle we scan over $\theta_{BL}=[0-0.3]$, the maximal value corresponds to the LHC bound on the Higgs scalar mixing angle \cite{Ilnicka:2018def}. The solid black line shows current experimental bounds from Xenon-1T \cite{Aprile:2017iyp}, the dashed black line shows the projected sensitivity for Xenon-nT \cite{Aprile:2015uzo} and the dotted black line shows the coherent neutrino scattering limit \cite{Billard:2013qya}.
}
\label{fig:mdm_mzbl}
\end{figure}

Having fixed $n_\chi=\!1/3$, we perform a random scan on the remaining six parameters in the model. In Fig.~\ref{fig:mdm_mzbl} we present our results, in the left panel we show the points in the $M_\chi-M_{Z_{BL}}$ plane that are in agreement with the measured relic abundance, $\Omega_\chi h^2 =0.1200 \pm 0.0012$~\cite{Aghanim:2018eyx}. All points shown satisfy bounds from direct detection and LEP. In the right panel, we show the predictions for the direct detection spin-independent cross-section as a function of the dark matter mass, for the same points as in the left panel. Red points correspond to $g_{BL}=0-0.25$, orange points correspond to $g_{BL}=0.25-0.5$, blue points correspond to $g_{BL}=0.5-0.75$ and green points correspond to $g_{BL}=0.75-\sqrt{\frac{\pi}{2}}$. The solid black line shows current experimental bounds from Xenon-1T \cite{Aprile:2017iyp}, the dashed black line shows the projected sensitivity for Xenon-nT \cite{Aprile:2015uzo} and the dotted black line shows the coherent neutrino scattering limit \cite{Billard:2013qya}. Since we know the maximal value for the gauge boson mass, we can show the predictions for the direct detection cross-section in the full parameter space. As can be seen, the Xenon-nT experiment will be able to probe a large fraction of the parameter space. In this model with $n_\chi=1/3$, we find the following upper bounds on the masses of the $B-L$ gauge boson and the dark matter candidate
$$M_{Z_{BL}} \lesssim 21 \ {\rm{TeV}} \  {\rm{and}} \ M_{\chi} \lesssim 11 \ {\rm{TeV}}.$$
Therefore, we can hope to test the Type I seesaw mechanism for Majorana neutrinos and this simple theory for dark matter in the near future.
\newpage
\section{LEPTON NUMBER AS A LOCAL GAUGE SYMMETRY}
\label{sec:U1L}
There are two simple gauge theories based on $\U(1)_L$ where one predicts the 
existence of a dark matter candidate from anomaly cancellation~\cite{Duerr:2013dza,Perez:2014qfa}. 
In this context, the dark matter mass is defined by the $\U(1)_L$ symmetry breaking scale and, 
as we will demonstrate, the scale must be in the multi-TeV scale in order to satisfy the relic density constraints.  
In Ref.~\cite{Duerr:2013dza}, it has been shown that one can cancel the anomalies by adding six new representations to the SM fermionic content plus the three right-handed neutrinos, and in this context the dark matter candidate can be either a Dirac or a Majorana, while in Ref.~\cite{Perez:2014qfa} it is shown that the theory can be anomaly-free by adding only four representations and the dark matter is predicted to always be a Majorana fermion. Since the main goal of this article is to investigate the most generic properties 
of a dark matter candidate in these theories, we will focus on the Majorana case and show the predictions 
in the context of a simplified model which describes the most important properties. 
Studies where the dark matter candidate is directly coupled only to leptons have been performed in Refs.~\cite{Fox:2008kb,
Kopp:2009et,
Bell:2014tta,
Freitas:2014jla,
delAguila:2014soa,
Chen:2015tia,
DEramo:2017zqw,
Madge:2018gfl}.

\subsection{Leptophilic Dark Matter}
\label{sec:leptophilic}
We consider a simple model for leptophilic Majorana dark matter which can be obtained in the context of the anomaly-free 
theories proposed in Refs.~\cite{Duerr:2013dza,Perez:2014qfa}. In this context, one has the SM leptons and the right-handed neutrinos
$$\ell_L \sim (1,2,-1/2,1), \hspace{0.3cm} e_R \sim (1,1,-1,1), \hspace{0.3cm} \nu_R \sim (1,1,0,1).$$
The new Higgs needed for spontaneous symmetry breaking $S_L \sim (1,1,0,3)$, with the leptonic charge fixed by anomaly cancellation, the dark matter candidate $\chi_L \sim (1,1,0,-3/2)$ 
and other fields needed for anomaly cancellation. The explicit extended fermionic sector for the UV completions of this simplified model has been relegated to Appendix~\ref{sec:AppendixA}. For details see Refs.~\cite{Duerr:2013dza,Perez:2014qfa}. 

The relevant Lagrangian for our discussions is given by
\begin{eqnarray}
\cal{L} &\supset & i \bar{\chi}_L \gamma^\mu D_\mu \chi_L \ + \  (D_\mu S_L)^\dagger (D^\mu S_L) - \left( \frac{y_\chi}{\sqrt{2}} \chi_L^T C \chi_L S_L \ + \ \rm{h.c.} \right), 
\end{eqnarray}
and, after spontaneous symmetry breaking, one finds the following physical interactions

\begin{figure}[tbp]
\centering
\includegraphics[width=0.48\linewidth]{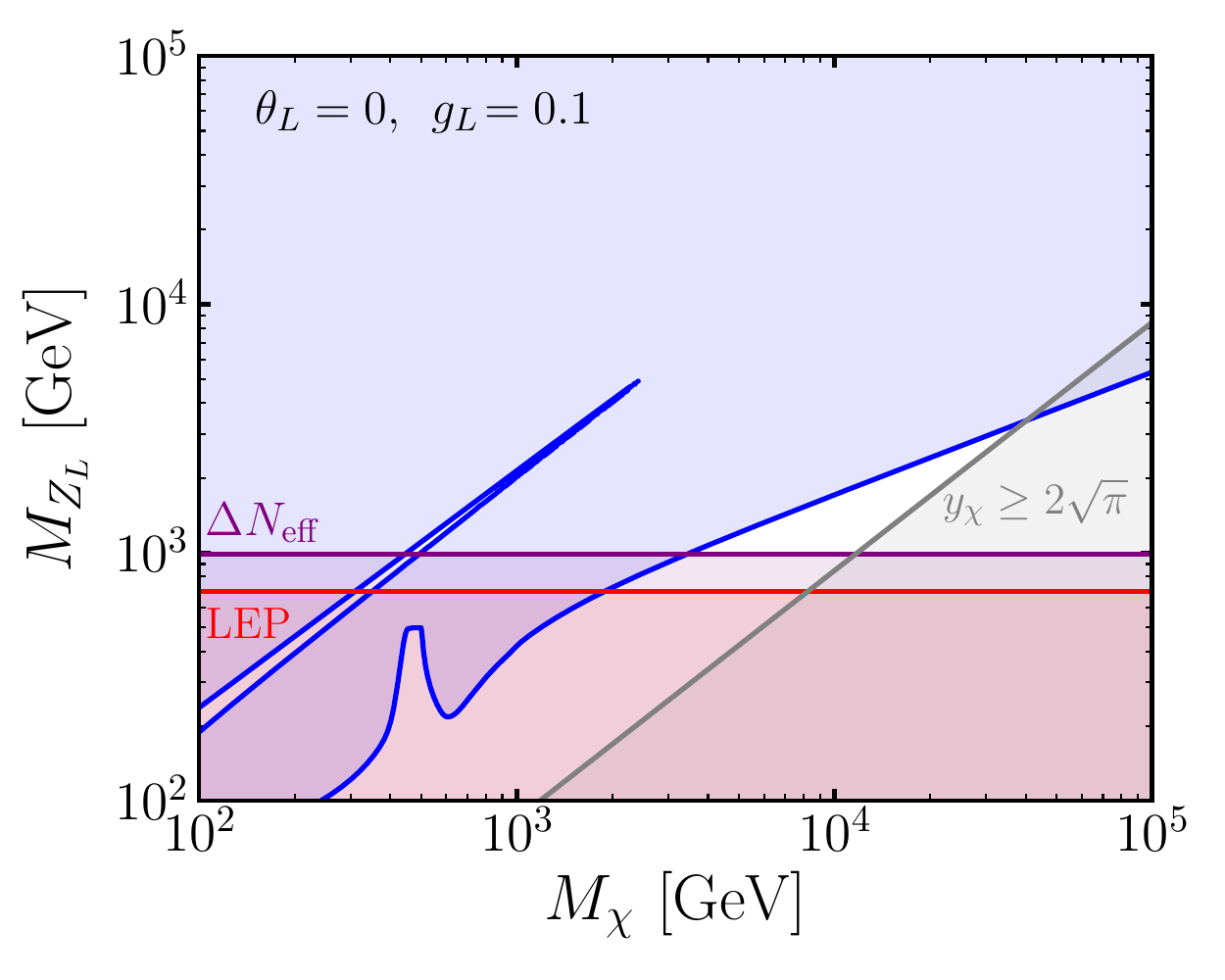} 
\includegraphics[width=0.48\linewidth]{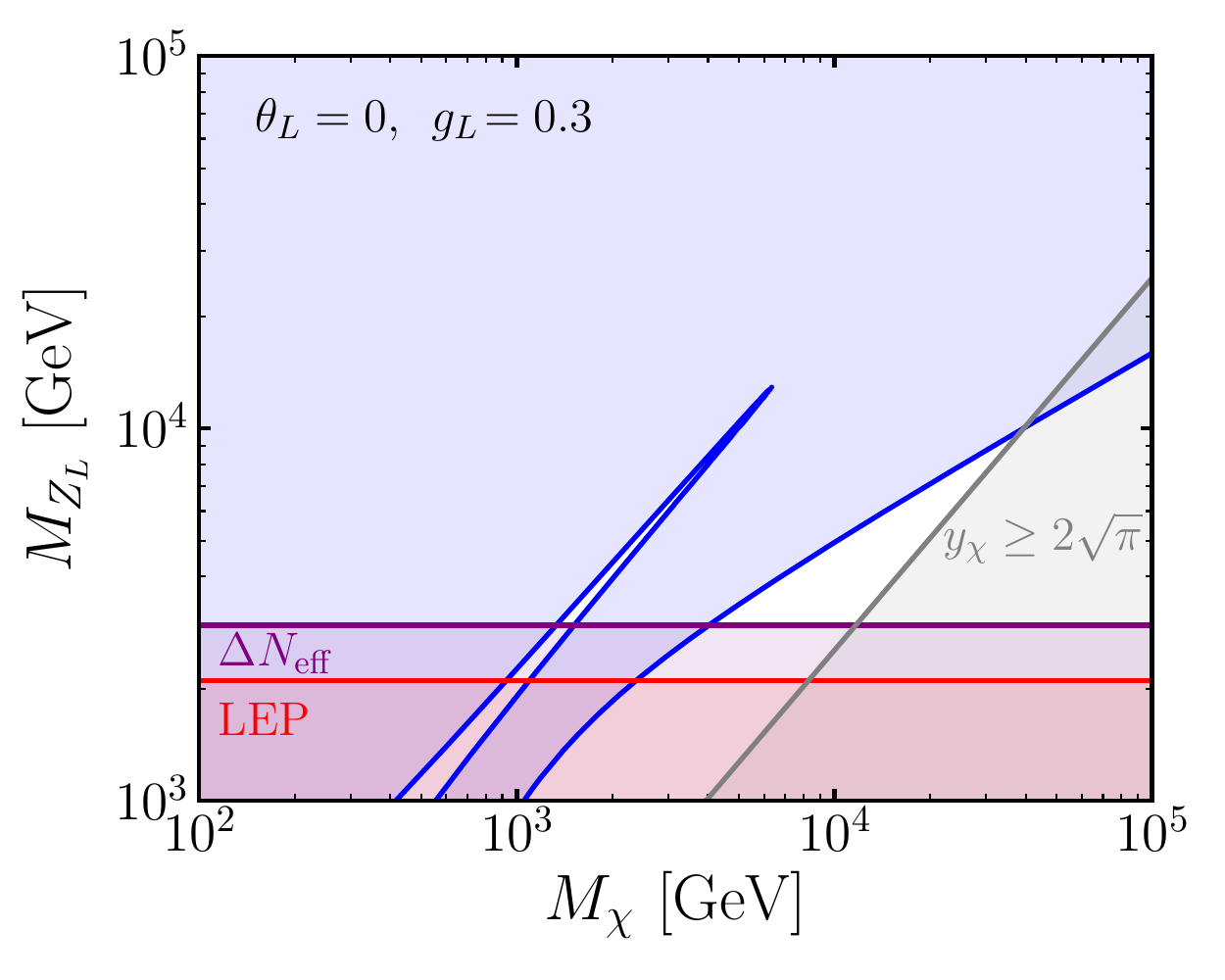} 
\includegraphics[width=0.48\linewidth]{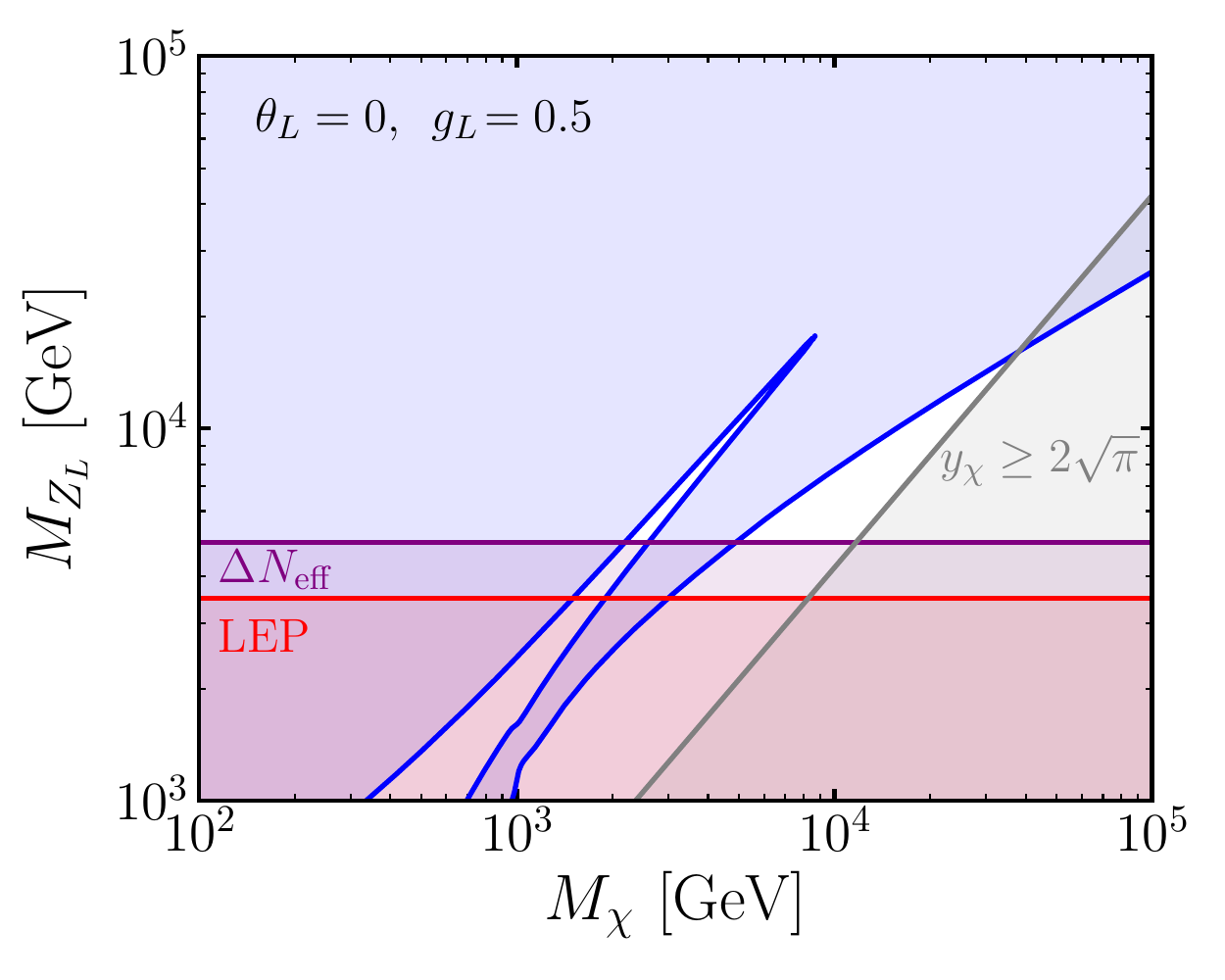} 
\includegraphics[width=0.48\linewidth]{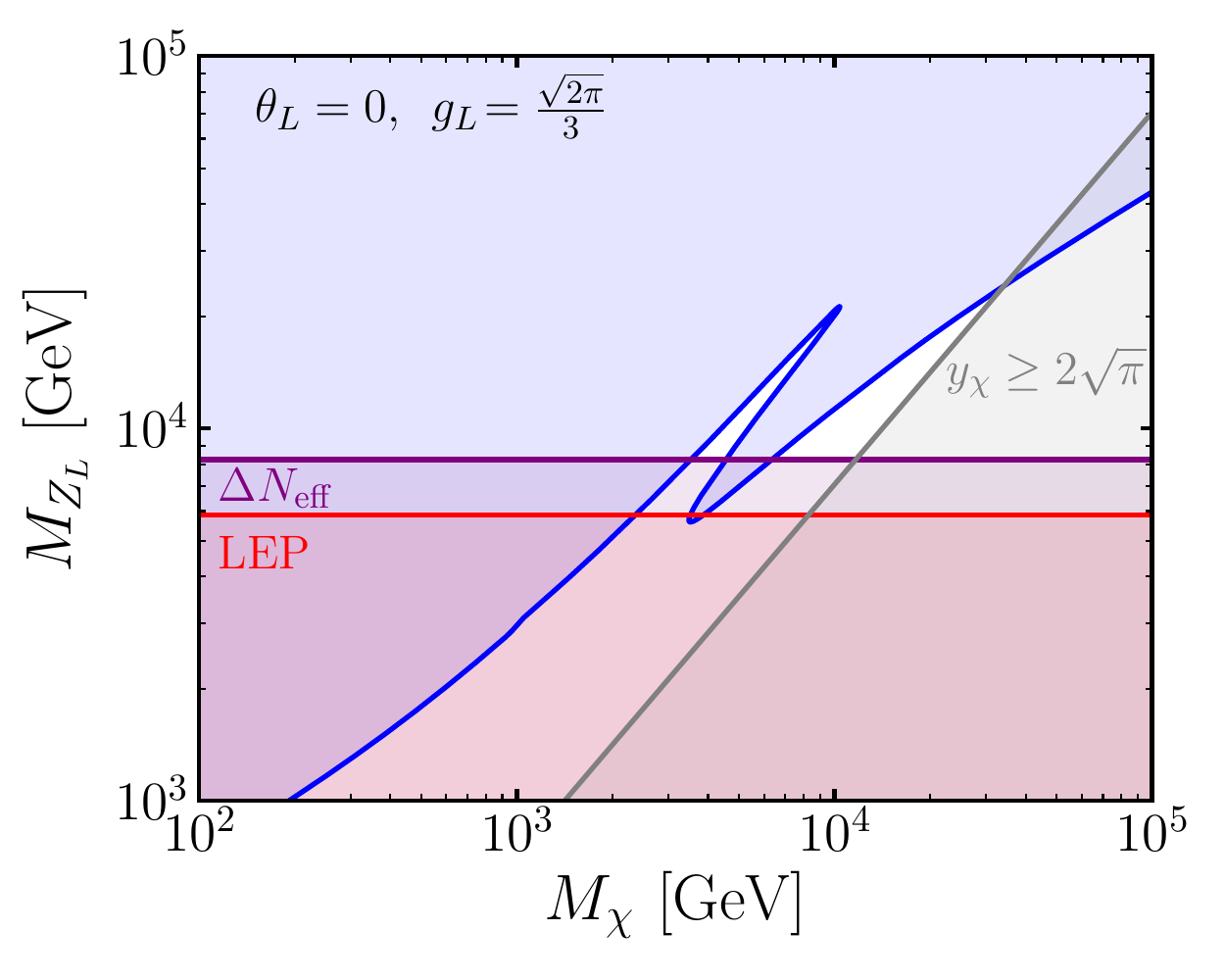} 
\caption{Results for the dark matter relic density in four different scenarios, $g_L=0.1$ (top-left), $g_L=0.3$ (top-right), $g_L=0.5$ (bottom-left) 
and $g_L=\sqrt{2 \pi}/3$ (bottom-right). We take $M_{h_2}=1$ TeV and no mixing angle. The solid blue line gives the measured dark matter relic density $\Omega_\chi h^2 =0.1200 \pm 0.0012$~\cite{Aghanim:2018eyx}, while the region shaded in blue overproduces it. The region shaded in gray is excluded by the perturbative bound on the Yukawa coupling $y_\chi$. 
The horizontal red band corresponds to the LEP~\cite{Alioli:2017nzr} bound on the $\U(1)_L$ gauge boson mass. The bounds from $N_{\rm eff}$ are shown by the purple region and rule out a large fraction of the parameter space.}
\label{fig:Relic-scenarios}
\end{figure}
\begin{eqnarray}
{\cal{L}} &\supset& \frac{3}{2} g_L \bar{\chi} \gamma^\mu \gamma^5 \chi Z^L_\mu - g_L \bar{\ell} \gamma^\mu  \ell Z^L_\mu - y_i \, \bar{\chi} \chi h_i - \frac{1}{2} M_\chi \chi^T C \chi, 
\end{eqnarray}
where $\ell=\nu_i,e_i$, with $i=1,2,3$, and $\chi=\chi^C$. In theories where the dark matter candidate is predicted by anomaly cancellation, the dark matter acquires mass through the mechanism of spontaneous symmetry breaking. This connection has phenomenological implications that will be discussed below. The Yukawa couplings in the above equation are given by
\begin{equation}
y_1 = \frac{M_\chi}{2 v_L} \sin \theta_L, \quad {\rm{and}}  \quad  y_2 = - \frac{M_\chi}{2 v_L} \cos \theta_L,
\end{equation}
and $M_{Z_L}=3 g_L v_L$. Thus, the gauge boson mass, $M_{Z_{L}}$, and the dark matter mass, $M_\chi$, are defined by the same symmetry breaking scale $v_L$. This model contains five free parameters:
\beq
M_\chi, \,\,\,\, M_{Z_{L}}, \,\,\,\, M_{h_2}, \,\,\,\, \theta_{L}, \,\,\,\, \text{and}  \,\,\,\, g_{L}.
\eeq
In this scenario, the dark matter charge $n_\chi$ is predicted by the theory. The relevant annihilation channels of our DM candidate in this theory are
$${\chi} \chi \to e_i^+ e_i^-, \, \bar{\nu}_i \nu_i, \, Z_{L} Z_{L}, \ Z_{L} h_i, \ h_i h_j, \ WW, \ ZZ, $$ 
where $h_i=h_1,h_2$ are the Higgses present in the theory; see Appendix~\ref{sec:AppendixB} for their explicit representation in Feynman graphs. In this model, the perturbative bound on the gauge coupling comes from the 
$S_{L}^\dagger S_{L} Z_{L} Z_{L}$ coupling and reads as $g_{L} \leq \sqrt{2 \pi}/3$.

In Fig.~\ref{fig:Relic-scenarios}, we present our result for the DM relic density and different constraints. The solid blue line saturates the relic abundance, $\Omega_\chi h^2 =0.1200 \pm 0.0012$~\cite{Aghanim:2018eyx}, and the region shaded in blue overproduces it. 
The four plots have two distinct regions where the correct relic abundance is achieved.  One corresponds to the resonance $M_\chi \approx M_{Z_L}/2$ in which annihilation into SM leptons gives the dominant contribution. The second one, to the right of the resonance, corresponds to the non-resonant region in which the annihilation channel $\chi \chi \rightarrow Z_L h_2$ gives the dominant contribution to the relic density.

The appearance of the non-resonant region arises due to the Yukawa interaction with the scalar $h_2$, and hence, new processes contribute to the dark matter annihilation channels, see Appendix~\ref{sec:AppendixB}. 
The area shaded in gray in Fig.~\ref{fig:Relic-scenarios} shows the excluded parameter space by the perturbative bound on the Yukawa coupling $y_\chi$. This bound gives an upper bound on the dark matter mass which, due to the connection with $M_{Z_L}$, translates as an upper bound on the lepton number breaking scale.

In this scenario, the new gauge boson is coupled only to leptons at tree-level, and hence, the LHC bounds given in Fig.~\ref{fig:Collider_bounds} cannot be applied. Even though  the coupling to quarks can be generated at the one-loop level, for a study including one-loop effects see \cite{DEramo:2017zqw}, and the bounds are weaker than the one coming from LEP~\cite{Alioli:2017nzr}. The latter are shown by the solid red line.

\begin{figure}[tbp]
\centering
\includegraphics[width=.6\textwidth]{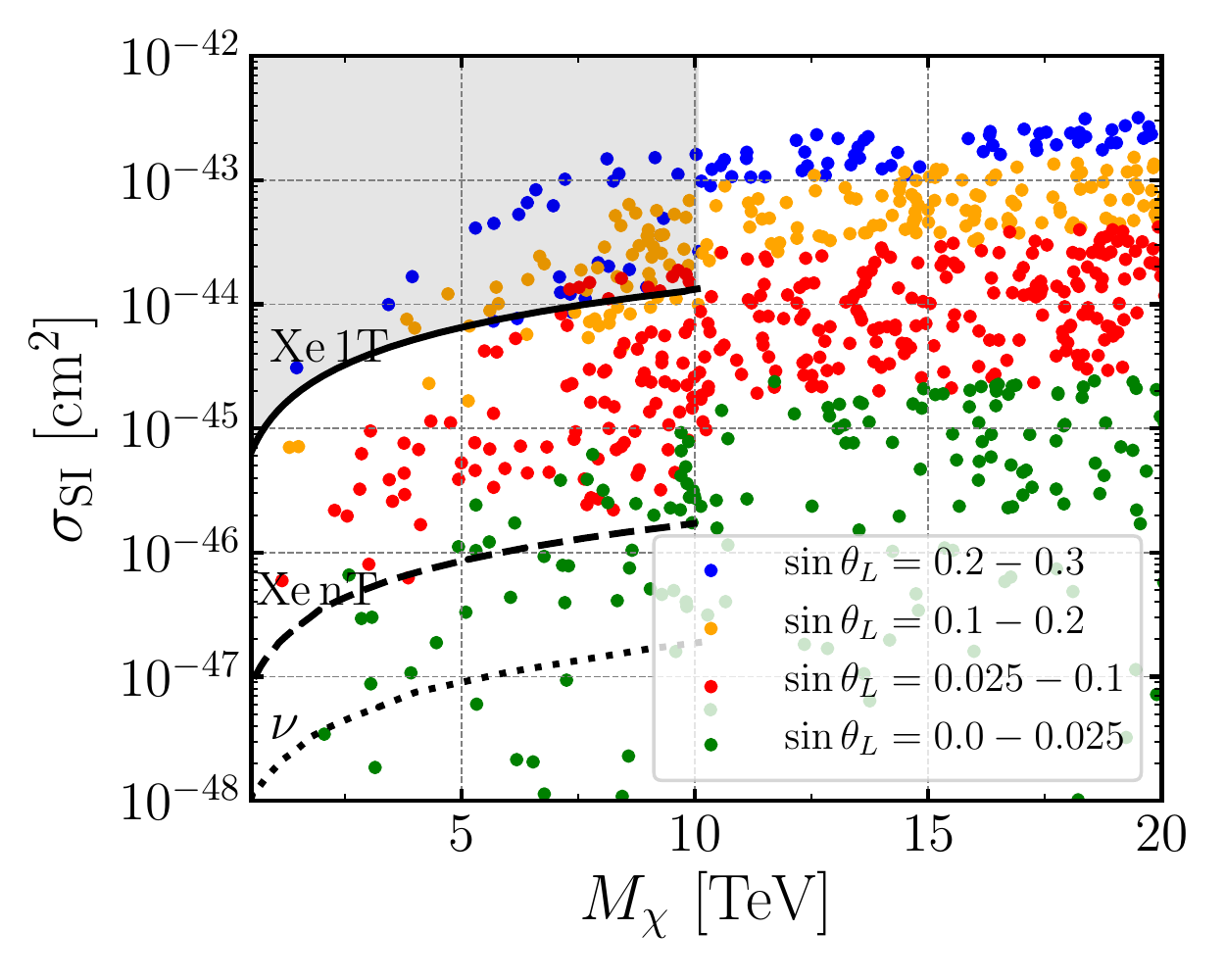} 
\caption{ 
Predictions for the direct detection spin-independent cross-section as a function of the dark matter mass in the $\U(1)_{L}$ scenario. We scan over $M_{h_2}=0.1-20$ TeV and apply the LHC bound~\cite{Aaboud:2017buh} on the mixing angle $\sin\theta_L\leq 0.3$. All points shown here satisfy the correct relic density $\Omega_\chi h^2 =0.1200 \pm 0.0036$. Green points correspond to $\sin \theta_L=0-0.025$, red points correspond to $\sin \theta_L =0.025-0.1$, orange points correspond to $\sin \theta_L =0.1-0.2$, and blue points correspond to $\sin \theta_L =0.2-0.3$. The solid black line shows current experimental bounds from Xenon-1T \cite{Aprile:2017iyp}, the dashed black line shows the projected sensitivity for Xenon-nT \cite{Aprile:2015uzo} and the dotted black line shows the coherent neutrino scattering limit~\cite{Billard:2013qya}.
}
\label{fig:DD}
\end{figure}

In the minimal model with one extra Higgs, the active neutrinos in the SM are predicted to be Dirac fermions. Therefore, as has been discussed in Section~\ref{sec:Neff}, there is a bound coming from the CMB measurement of the effective number of neutrino species $N_{\rm eff}$, which gives the following bound,
\beq
\Delta N_{\rm eff} < 0.285 \hspace{0.5cm} \Rightarrow \hspace{0.5cm} \frac{M_{Z_L}}{g_L} > 9.87 \,\, {\rm TeV}.
\eeq
Similarly to the Stueckelberg case, it is stronger than the LEP bound and it is shown by the solid purple line in Fig.\ref{fig:Relic-scenarios}. We should stress that by adding a new Higgs scalar with lepton number $L = 2$, a Majorana mass term can be written for $\nu_R$ and, if these states are heavy, the bound from $N_{\rm eff}$ is not relevant.

The upper left panel in Fig.~\ref{fig:Relic-scenarios} corresponds to $g_L=0.1$. In this case, the bound from $\Delta N_{\rm eff}$ requires $M_{Z_L}>0.99$ TeV and the correct relic abundance can be produced close to the resonance $M_\chi \approx M_{Z_L}/2$ for dark matter masses $M_\chi \approx 550 \,\, {\rm GeV} - 2.4$ TeV. In the non-resonant regime, the correct relic abundance can be generated for dark matter masses $M_\chi \approx 5.4 - 39$ TeV. For larger dark matter masses, the Yukawa coupling, $y_\chi$, becomes non-perturbative.
In the lower right panel of Fig.~\ref{fig:Relic-scenarios}, we show our results for the maximal value of $g_L$ allowed by perturbativity, $g_L = \sqrt{2\pi}/3 \approx 0.84$. The bound from $\Delta N_{\rm eff}$ requires $M_{Z_L}\!>\!8.25$ TeV. The resonant region that saturates the relic density and satisfies the bounds from LEP and $\Delta N_{\rm eff}$ corresponds to $M_\chi \approx 4.6 \,\, {\rm GeV} - 10.2$ TeV, while the non-resonant region works for dark matter masses $M_\chi \approx 8.9 - 34$ TeV. Above this value, the Yukawa coupling becomes non-perturbative.
Therefore, the upper bounds correspond to $M_{Z_L} \lesssim 21$ TeV and $ M_\chi \lesssim 34$  TeV. 

Regarding direct detection, the $\chi\!-\!N$ interaction can be mediated by Higgs mixing or the exchange of a $Z_L$. The latter is not coupled to quarks at tree-level, and hence, this process is loop suppressed \citep{DEramo:2017zqw}. Moreover, due to the Majorana nature of dark matter, there will also be velocity suppression. Hence, we focus on the contribution from Higgs mixing,
\beq
\sigma_{\chi N}^\text{SI}(h_i)=\frac{72 G_F}{\sqrt{2} 4 \pi}\sin^2 \theta_L \cos^2 \theta_L m_N^4 \frac{g_L^2 M_\chi^2}{M_{Z_L}^2}\left(\frac{1}{M_{h_1}^2}-\frac{1}{M_{h_2}^2}\right)^2 f_N^2,
\eeq
where $m_N$ corresponds to the nucleon mass, $G_F$ is the Fermi constant, and for the effective Higgs-nucleon-nucleon coupling we take $f_N=0.3$ \cite{Alarcon:2011zs, Hoferichter:2017olk}.

In Fig.~\ref{fig:DD}, we present our predictions for the spin-independent cross-section as a function of the dark matter mass in the $\U(1)_{L}$ scenario. We perform a scan over $M_{h_2}=0.1-20$ TeV and apply the LHC bound on the scalar mixing angle $\sin \theta_L \leq0.3$ \cite{Ilnicka:2018def}. Green points correspond to $\sin \theta_L=0-0.025$, red points correspond to $\sin \theta_L=0.025-0.1$, orange points correspond to $\sin \theta_L=0.1-0.2$, and blue points correspond to $\sin \theta_L=0.2-0.3$. 
The solid black line shows current experimental bounds from Xenon-1T \cite{Aprile:2017iyp}, the dashed black line shows the projected sensitivity for Xenon-nT \cite{Aprile:2015uzo}, and the dotted black line shows the coherent neutrino scattering limit~\cite{Billard:2013qya}.

As illustrated in Fig.~\ref{fig:DD}, for dark matter masses within reach of Xenon-nT, $M_\chi<10$ TeV, this type of experiments will be able to probe scalar mixing angles $\theta_L>0.025$, providing a stronger constraint than colliders on the Higgs mixing angle.
In this scenario, there is no strong correlation between the $\Delta N_{\rm eff}$ and direct detection bounds because the main contribution to direct detection is mediated by the Higgses in the theory, 
while $\Delta N_{\rm eff}$ provides a constraint on the ratio $M_{Z_L}/g_L$.

%

\section{SUMMARY}
\label{sec:Summary}
In this work, we investigated possible connections between the origin of neutrino masses and the properties of dark matter candidates in simple gauge theories 
based on local $B\!-\!L$ or $L$ symmetries. In theories based on $B\!-\!L$, the gauge boson mass can be generated through the 
Stueckelberg or the Higgs mechanism. In the Canonical Seesaw scenario, the $B\!-\!L$ symmetry is spontaneously broken in two units via 
the Higgs mechanism, the neutrinos are Majorana fermions and, in the simplest model, the dark matter is a Dirac fermion.
In the case of anomaly-free gauge theories based on $\U(1)_L$, the Higgs mechanism is needed to generate the masses for the new fermions present in the theory for anomaly cancellation.
We studied the simplest theories for local lepton number where the existence of a dark matter candidate is predicted from anomaly cancellation. In this case, the lepton number is broken in 
three units and the neutrinos are predicted to be Dirac particles, while the dark matter candidate is a Majorana fermion in most generic models. 

We showed that the cosmological constraint on the relic dark matter density implies that the upper bound on the symmetry breaking scale in these theories, where a connection between the origin of neutrino masses and the dark matter can be made, is in the multi-TeV region. In addition, we demonstrated that in theories where the neutrinos are Dirac, namely, the $B\!-\!L$ Stueckelberg scenario and the theory based on $\U(1)_L$, the cosmological bound on the effective number of neutrino species, $\Delta N_{\rm eff}$, provides a strong bound in the parameter space of the models. Furthermore, the projected sensitivity to this parameter by the CMB Stage-IV experiments could fully probe the parameter space that also explains dark matter. These results allow us to be optimistic about the testability of the mechanism for neutrino masses in current and future experiments.

\vspace{1.0cm}

{\textit{Acknowledgments}: We would like to thank the organizers of PHENO2019 for this nice meeting where this article was completed. 
We thank Yue Zhang for discussions. The work of C.M. has been supported in part by Grants No. FPA2014-53631-C2-1-P, FPA2017-84445-P, and SEV-2014-0398 (AEI/ERDF, EU), and the  La Caixa-Severo Ochoa scholarship.}

\newpage
\appendix

\section{Anomaly-free $U(1)_L$ theories}
\label{sec:AppendixA}

The needed fermionic representations to define anomaly-free theories based on $\U(1)_L$ are listed in the tables below.
In Table 1 we have the extra fermions in the model proposed in Ref.~\cite{Perez:2014qfa}, while in Table 2 we show 
the representations needed in Ref.~\cite{Duerr:2013dza}.

\vspace{1.0cm}

\begin{table}[h]\setlength{\bigstrutjot}{6pt}
\centering
\begin{tabular}{|ccccc|}\hline
    Fields & $\SU(3)_C$ & $\SU(2)_L$ & $\U(1)_Y$  & $\U(1)_L$ \bigstrut\\\hline\hline
    $ \Psi_L = \mqty(\Psi_L^+ \\ \Psi_L^0)$ & ${1}$ & ${2}$ & $\frac{1}{2}$ & $\frac{3}{2}$\bigstrut\\
    $ \Psi_R = \mqty(\Psi_R^+ \\ \Psi_R^0)$ & $1$ & $2$ & $\frac{1}{2}$ & $-\frac{3}{2}$  \bigstrut\\
    $\Sigma_L = \frac{1}{\sqrt{2}} \mqty(\Sigma^0_L & \sqrt{2}\Sigma^+_L \\ \sqrt{2}\Sigma^-_L & -\Sigma^0_L)$ & $1$ & $3$ & $0$ & $-\frac{3}{2}$ \bigstrut\\
   $\chi_L^0$ & $1$ & $1$ & $0$ & $-\frac{3}{2}$ \bigstrut\\ \hline
  \end{tabular}
  \caption{Fermionic representations in the model proposed in Ref.~\cite{Perez:2014qfa}.}
  \label{tabla2}
\end{table}%

\vspace{1.0cm}

\begin{table}[h]\setlength{\bigstrutjot}{6pt}
\centering
\begin{tabular}{|ccccc|}\hline
    Fields & $\SU(3)_C$ & $\SU(2)_L$ & $\U(1)_Y$  & $\U(1)_L$ \bigstrut\\\hline\hline
    $\Psi_L = \mqty(\Psi_L^0 \\ \Psi_L^-)$ & ${1}$ & ${2}$ & $-\frac{1}{2}$ & $-\frac{3}{2}$\bigstrut\\
    $\Psi_R = \mqty(\Psi_R^0 \\ \Psi_R^-)$ & $1$ & $2$ & $-\frac{1}{2}$ & $\frac{3}{2}$  \bigstrut\\
    $\eta_R^-$ & $ 1$ & $ 1$ & $-1$ & $-\frac{3}{2}$ \bigstrut\\
    $\eta_L^-$ & $1$ & $1$ & $-1$ & $\frac{3}{2}$ \bigstrut\\
    $\chi_R^0$ & $1$ & $1$ & $0$ & $-\frac{3}{2}$ \bigstrut\\
    $\chi_L^0$ & $1$ & $1$ & $0$ & $\frac{3}{2}$ \bigstrut\\
\hline
  \end{tabular}
  \caption{Fermionic representations in the model proposed in Ref.~\cite{Duerr:2013dza}.}
  \label{tabla1}
\end{table}

\newpage

\section{Feynman diagrams for dark matter annihilation}
\label{sec:AppendixB}

\begin{figure}[h]
\centering
\includegraphics[width=.75\textwidth]{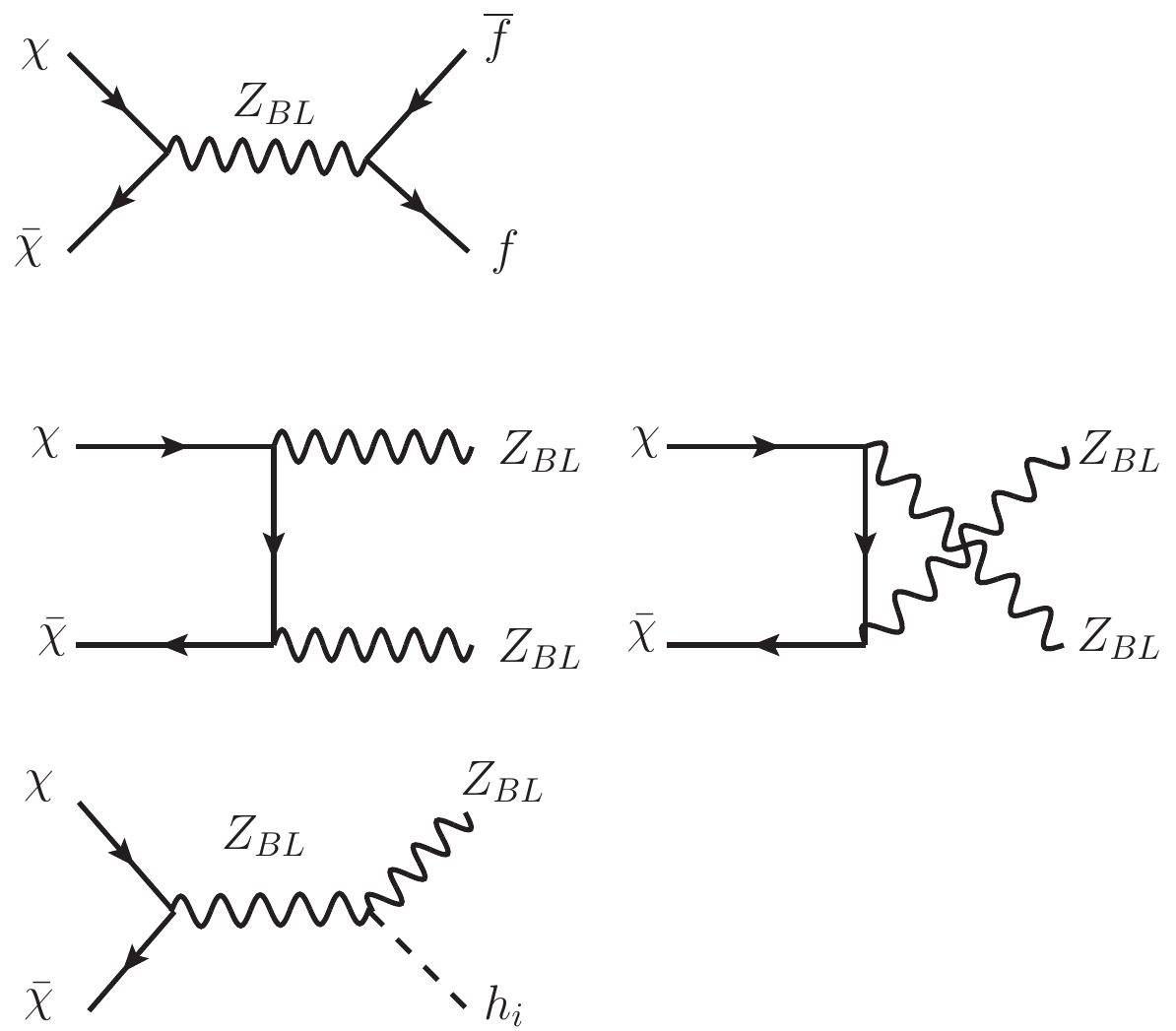} 
\caption{Feynman diagrams for the dark matter annihilation channels in the  $\U(1)_{B-L}$ theories. 
}
\label{fig:Diagrams_BL}
\end{figure}

\begin{figure}[t]
\centering
\includegraphics[width=1.0\textwidth]{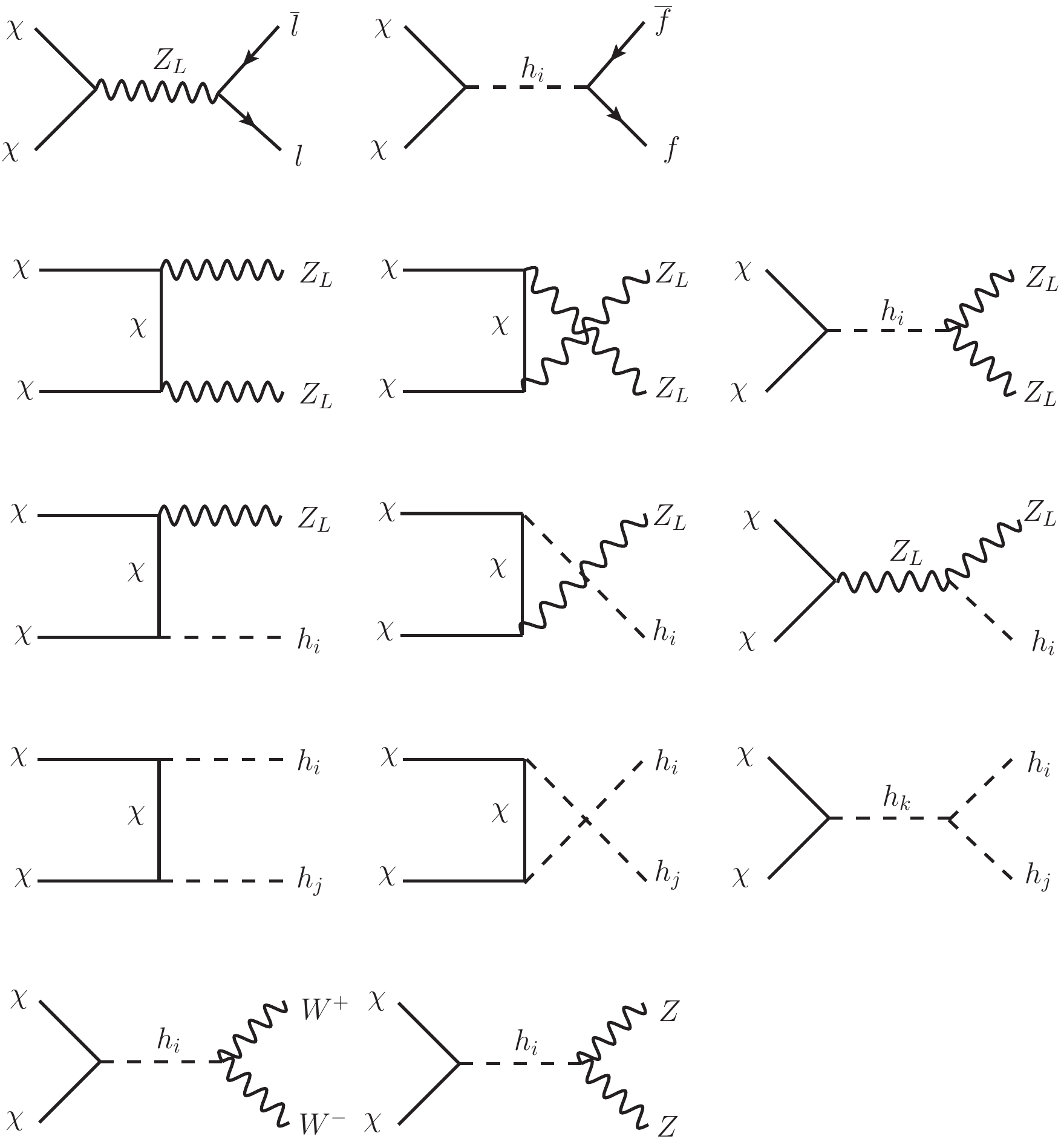} 
\caption{Feynman diagrams for the dark matter annihilation channels in the  $\U(1)_L$ theories.
}
\label{fig:Diagrams_L}
\end{figure}

In this appendix we present the Feynman diagrams for the dark matter annihilation channels. The diagrams shown in Fig.~\ref{fig:Diagrams_BL} correspond to the $\U(1)_{B-L}$ case. For the Stueckelberg scenario the diagrams involving a Higgs are not taken into account. The diagrams in Fig.~\ref{fig:Diagrams_L} correspond to the scenario with $\U(1)_L$.

\newpage
\bibliography{Nu-DM}

\begin{thebibliography}{72}%
\makeatletter
\providecommand \@ifxundefined [1]{%
 \@ifx{#1\undefined}
}%
\providecommand \@ifnum [1]{%
 \ifnum #1\expandafter \@firstoftwo
 \else \expandafter \@secondoftwo
 \fi
}%
\providecommand \@ifx [1]{%
 \ifx #1\expandafter \@firstoftwo
 \else \expandafter \@secondoftwo
 \fi
}%
\providecommand \natexlab [1]{#1}%
\providecommand \enquote  [1]{``#1''}%
\providecommand \bibnamefont  [1]{#1}%
\providecommand \bibfnamefont [1]{#1}%
\providecommand \citenamefont [1]{#1}%
\providecommand \href@noop [0]{\@secondoftwo}%
\providecommand \href [0]{\begingroup \@sanitize@url \@href}%
\providecommand \@href[1]{\@@startlink{#1}\@@href}%
\providecommand \@@href[1]{\endgroup#1\@@endlink}%
\providecommand \@sanitize@url [0]{\catcode `\\12\catcode `\$12\catcode
  `\&12\catcode `\#12\catcode `\^12\catcode `\_12\catcode `\%12\relax}%
\providecommand \@@startlink[1]{}%
\providecommand \@@endlink[0]{}%
\providecommand \url  [0]{\begingroup\@sanitize@url \@url }%
\providecommand \@url [1]{\endgroup\@href {#1}{\urlprefix }}%
\providecommand \urlprefix  [0]{URL }%
\providecommand \Eprint [0]{\href }%
\providecommand \doibase [0]{http://dx.doi.org/}%
\providecommand \selectlanguage [0]{\@gobble}%
\providecommand \bibinfo  [0]{\@secondoftwo}%
\providecommand \bibfield  [0]{\@secondoftwo}%
\providecommand \translation [1]{[#1]}%
\providecommand \BibitemOpen [0]{}%
\providecommand \bibitemStop [0]{}%
\providecommand \bibitemNoStop [0]{.\EOS\space}%
\providecommand \EOS [0]{\spacefactor3000\relax}%
\providecommand \BibitemShut  [1]{\csname bibitem#1\endcsname}%
\let\auto@bib@innerbib\@empty
\bibitem [{\citenamefont {Esteban}\ \emph {et~al.}(2019)\citenamefont
  {Esteban}, \citenamefont {Gonzalez-Garcia}, \citenamefont
  {Hernandez-Cabezudo}, \citenamefont {Maltoni},\ and\ \citenamefont
  {Schwetz}}]{Esteban:2018azc}%
  \BibitemOpen
  \bibfield  {author} {\bibinfo {author} {\bibfnamefont {I.}~\bibnamefont
  {Esteban}}, \bibinfo {author} {\bibfnamefont {M.~C.}\ \bibnamefont
  {Gonzalez-Garcia}}, \bibinfo {author} {\bibfnamefont {A.}~\bibnamefont
  {Hernandez-Cabezudo}}, \bibinfo {author} {\bibfnamefont {M.}~\bibnamefont
  {Maltoni}}, \ and\ \bibinfo {author} {\bibfnamefont {T.}~\bibnamefont
  {Schwetz}},\ }\bibfield  {title} {\enquote {\bibinfo {title} {{Global
  analysis of three-flavour neutrino oscillations: synergies and tensions in
  the determination of $\theta_23, \delta_CP$, and the mass ordering}},}\
  }\href {\doibase 10.1007/JHEP01(2019)106} {\bibfield  {journal} {\bibinfo
  {journal} {JHEP}\ }\textbf {\bibinfo {volume} {01}},\ \bibinfo {pages} {106}
  (\bibinfo {year} {2019})},\ \Eprint {http://arxiv.org/abs/1811.05487}
  {arXiv:1811.05487 [hep-ph]} \BibitemShut {NoStop}%
\bibitem [{\citenamefont {Minkowski}(1977)}]{Minkowski:1977sc}%
  \BibitemOpen
  \bibfield  {author} {\bibinfo {author} {\bibfnamefont {P.}~\bibnamefont
  {Minkowski}},\ }\bibfield  {title} {\enquote {\bibinfo {title} {{$\mu \to
  e\gamma$ at a Rate of One Out of $10^{9}$ Muon Decays?}}}\ }\href {\doibase
  10.1016/0370-2693(77)90435-X} {\bibfield  {journal} {\bibinfo  {journal}
  {Phys. Lett.}\ }\textbf {\bibinfo {volume} {67B}},\ \bibinfo {pages}
  {421--428} (\bibinfo {year} {1977})}\BibitemShut {NoStop}%
\bibitem [{\citenamefont {Gell-Mann}\ \emph {et~al.}(1979)\citenamefont
  {Gell-Mann}, \citenamefont {Ramond},\ and\ \citenamefont
  {Slansky}}]{GellMann:1980vs}%
  \BibitemOpen
  \bibfield  {author} {\bibinfo {author} {\bibfnamefont {M.}~\bibnamefont
  {Gell-Mann}}, \bibinfo {author} {\bibfnamefont {P.}~\bibnamefont {Ramond}}, \
  and\ \bibinfo {author} {\bibfnamefont {R.}~\bibnamefont {Slansky}},\
  }\bibfield  {title} {\enquote {\bibinfo {title} {{Complex Spinors and Unified
  Theories}},}\ }\bibfield  {booktitle} {\emph {\bibinfo {booktitle}
  {{Supergravity Workshop Stony Brook, New York, September 27-28, 1979}}},\
  }\href@noop {} {\bibfield  {journal} {\bibinfo  {journal} {Conf. Proc.}\
  }\textbf {\bibinfo {volume} {C790927}},\ \bibinfo {pages} {315--321}
  (\bibinfo {year} {1979})},\ \Eprint {http://arxiv.org/abs/1306.4669}
  {arXiv:1306.4669 [hep-th]} \BibitemShut {NoStop}%
\bibitem [{\citenamefont {Yanagida}(1979)}]{Yanagida:1979as}%
  \BibitemOpen
  \bibfield  {author} {\bibinfo {author} {\bibfnamefont {T.}~\bibnamefont
  {Yanagida}},\ }\bibfield  {title} {\enquote {\bibinfo {title} {{Horizontal
  gauge symmetry and masses of neutrinos}},}\ }\bibfield  {booktitle} {\emph
  {\bibinfo {booktitle} {{Proceedings: Workshop on the Unified Theories and the
  Baryon Number in the Universe: Tsukuba, Japan, February 13-14, 1979}}},\
  }\href@noop {} {\bibfield  {journal} {\bibinfo  {journal} {Conf. Proc.}\
  }\textbf {\bibinfo {volume} {C7902131}},\ \bibinfo {pages} {95--99} (\bibinfo
  {year} {1979})}\BibitemShut {NoStop}%
\bibitem [{\citenamefont {Mohapatra}\ and\ \citenamefont
  {Senjanovic}(1980)}]{Mohapatra:1979ia}%
  \BibitemOpen
  \bibfield  {author} {\bibinfo {author} {\bibfnamefont {R.~N.}\ \bibnamefont
  {Mohapatra}}\ and\ \bibinfo {author} {\bibfnamefont {G.}~\bibnamefont
  {Senjanovic}},\ }\bibfield  {title} {\enquote {\bibinfo {title} {{Neutrino
  Mass and Spontaneous Parity Nonconservation}},}\ }\href {\doibase
  10.1103/PhysRevLett.44.912} {\bibfield  {journal} {\bibinfo  {journal} {Phys.
  Rev. Lett.}\ }\textbf {\bibinfo {volume} {44}},\ \bibinfo {pages} {912}
  (\bibinfo {year} {1980})},\ \bibinfo {note} {[,231(1979)]}\BibitemShut
  {NoStop}%
\bibitem [{\citenamefont {Fileviez~Perez}\ and\ \citenamefont
  {Murgui}(2018)}]{FileviezPerez:2018toq}%
  \BibitemOpen
  \bibfield  {author} {\bibinfo {author} {\bibfnamefont {P.}~\bibnamefont
  {Fileviez~Perez}}\ and\ \bibinfo {author} {\bibfnamefont {C.}~\bibnamefont
  {Murgui}},\ }\bibfield  {title} {\enquote {\bibinfo {title} {{Dark Matter and
  The Seesaw Scale}},}\ }\href {\doibase 10.1103/PhysRevD.98.055008} {\bibfield
   {journal} {\bibinfo  {journal} {Phys. Rev.}\ }\textbf {\bibinfo {volume}
  {D98}},\ \bibinfo {pages} {055008} (\bibinfo {year} {2018})},\ \Eprint
  {http://arxiv.org/abs/1803.07462} {arXiv:1803.07462 [hep-ph]} \BibitemShut
  {NoStop}%
\bibitem [{\citenamefont {Okada}\ and\ \citenamefont
  {Seto}(2010)}]{Okada:2010wd}%
  \BibitemOpen
  \bibfield  {author} {\bibinfo {author} {\bibfnamefont {N.}~\bibnamefont
  {Okada}}\ and\ \bibinfo {author} {\bibfnamefont {O.}~\bibnamefont {Seto}},\
  }\bibfield  {title} {\enquote {\bibinfo {title} {{Higgs portal dark matter in
  the minimal gauged $U(1)_{B-L}$ model}},}\ }\href {\doibase
  10.1103/PhysRevD.82.023507} {\bibfield  {journal} {\bibinfo  {journal} {Phys.
  Rev.}\ }\textbf {\bibinfo {volume} {D82}},\ \bibinfo {pages} {023507}
  (\bibinfo {year} {2010})},\ \Eprint {http://arxiv.org/abs/1002.2525}
  {arXiv:1002.2525 [hep-ph]} \BibitemShut {NoStop}%
\bibitem [{\citenamefont {Basak}\ and\ \citenamefont
  {Mondal}(2014)}]{Basak:2013cga}%
  \BibitemOpen
  \bibfield  {author} {\bibinfo {author} {\bibfnamefont {T.}~\bibnamefont
  {Basak}}\ and\ \bibinfo {author} {\bibfnamefont {T.}~\bibnamefont {Mondal}},\
  }\bibfield  {title} {\enquote {\bibinfo {title} {{Constraining Minimal
  $U(1)_{B-L}$ model from Dark Matter Observations}},}\ }\href {\doibase
  10.1103/PhysRevD.89.063527} {\bibfield  {journal} {\bibinfo  {journal} {Phys.
  Rev.}\ }\textbf {\bibinfo {volume} {D89}},\ \bibinfo {pages} {063527}
  (\bibinfo {year} {2014})},\ \Eprint {http://arxiv.org/abs/1308.0023}
  {arXiv:1308.0023 [hep-ph]} \BibitemShut {NoStop}%
\bibitem [{\citenamefont {Kanemura}\ \emph {et~al.}(2014)\citenamefont
  {Kanemura}, \citenamefont {Matsui},\ and\ \citenamefont
  {Sugiyama}}]{Kanemura:2014rpa}%
  \BibitemOpen
  \bibfield  {author} {\bibinfo {author} {\bibfnamefont {S.}~\bibnamefont
  {Kanemura}}, \bibinfo {author} {\bibfnamefont {T.}~\bibnamefont {Matsui}}, \
  and\ \bibinfo {author} {\bibfnamefont {H.}~\bibnamefont {Sugiyama}},\
  }\bibfield  {title} {\enquote {\bibinfo {title} {{Neutrino mass and dark
  matter from gauged $U(1)_{B-L}$ breaking}},}\ }\href {\doibase
  10.1103/PhysRevD.90.013001} {\bibfield  {journal} {\bibinfo  {journal} {Phys.
  Rev.}\ }\textbf {\bibinfo {volume} {D90}},\ \bibinfo {pages} {013001}
  (\bibinfo {year} {2014})},\ \Eprint {http://arxiv.org/abs/1405.1935}
  {arXiv:1405.1935 [hep-ph]} \BibitemShut {NoStop}%
\bibitem [{\citenamefont {Duerr}\ \emph {et~al.}(2015)\citenamefont {Duerr},
  \citenamefont {Fileviez~Perez},\ and\ \citenamefont
  {Smirnov}}]{Duerr:2015wfa}%
  \BibitemOpen
  \bibfield  {author} {\bibinfo {author} {\bibfnamefont {M.}~\bibnamefont
  {Duerr}}, \bibinfo {author} {\bibfnamefont {P.}~\bibnamefont
  {Fileviez~Perez}}, \ and\ \bibinfo {author} {\bibfnamefont {J.}~\bibnamefont
  {Smirnov}},\ }\bibfield  {title} {\enquote {\bibinfo {title} {{Simplified
  Dirac Dark Matter Models and Gamma-Ray Lines}},}\ }\href {\doibase
  10.1103/PhysRevD.92.083521} {\bibfield  {journal} {\bibinfo  {journal} {Phys.
  Rev.}\ }\textbf {\bibinfo {volume} {D92}},\ \bibinfo {pages} {083521}
  (\bibinfo {year} {2015})},\ \Eprint {http://arxiv.org/abs/1506.05107}
  {arXiv:1506.05107 [hep-ph]} \BibitemShut {NoStop}%
\bibitem [{\citenamefont {Ma}\ \emph {et~al.}(2015)\citenamefont {Ma},
  \citenamefont {Pollard}, \citenamefont {Srivastava},\ and\ \citenamefont
  {Zakeri}}]{Ma:2015mjd}%
  \BibitemOpen
  \bibfield  {author} {\bibinfo {author} {\bibfnamefont {E.}~\bibnamefont
  {Ma}}, \bibinfo {author} {\bibfnamefont {N.}~\bibnamefont {Pollard}},
  \bibinfo {author} {\bibfnamefont {R.}~\bibnamefont {Srivastava}}, \ and\
  \bibinfo {author} {\bibfnamefont {M.}~\bibnamefont {Zakeri}},\ }\bibfield
  {title} {\enquote {\bibinfo {title} {{Gauge $B-L$ Model with Residual $Z_3$
  Symmetry}},}\ }\href {\doibase 10.1016/j.physletb.2015.09.010} {\bibfield
  {journal} {\bibinfo  {journal} {Phys. Lett.}\ }\textbf {\bibinfo {volume}
  {B750}},\ \bibinfo {pages} {135--138} (\bibinfo {year} {2015})},\ \Eprint
  {http://arxiv.org/abs/1507.03943} {arXiv:1507.03943 [hep-ph]} \BibitemShut
  {NoStop}%
\bibitem [{\citenamefont {Wang}\ and\ \citenamefont
  {Han}(2015)}]{Wang:2015saa}%
  \BibitemOpen
  \bibfield  {author} {\bibinfo {author} {\bibfnamefont {W.}~\bibnamefont
  {Wang}}\ and\ \bibinfo {author} {\bibfnamefont {Z.-L.}\ \bibnamefont {Han}},\
  }\bibfield  {title} {\enquote {\bibinfo {title} {{Radiative linear seesaw
  model, dark matter, and $U(1)_{B-L}$}},}\ }\href {\doibase
  10.1103/PhysRevD.92.095001} {\bibfield  {journal} {\bibinfo  {journal} {Phys.
  Rev.}\ }\textbf {\bibinfo {volume} {D92}},\ \bibinfo {pages} {095001}
  (\bibinfo {year} {2015})},\ \Eprint {http://arxiv.org/abs/1508.00706}
  {arXiv:1508.00706 [hep-ph]} \BibitemShut {NoStop}%
\bibitem [{\citenamefont {Okada}\ and\ \citenamefont
  {Okada}(2016)}]{Okada:2016gsh}%
  \BibitemOpen
  \bibfield  {author} {\bibinfo {author} {\bibfnamefont {N.}~\bibnamefont
  {Okada}}\ and\ \bibinfo {author} {\bibfnamefont {S.}~\bibnamefont {Okada}},\
  }\bibfield  {title} {\enquote {\bibinfo {title} {{$Z^\prime_{BL}$ portal dark
  matter and LHC Run-2 results}},}\ }\href {\doibase
  10.1103/PhysRevD.93.075003} {\bibfield  {journal} {\bibinfo  {journal} {Phys.
  Rev.}\ }\textbf {\bibinfo {volume} {D93}},\ \bibinfo {pages} {075003}
  (\bibinfo {year} {2016})},\ \Eprint {http://arxiv.org/abs/1601.07526}
  {arXiv:1601.07526 [hep-ph]} \BibitemShut {NoStop}%
\bibitem [{\citenamefont {Kaneta}\ \emph {et~al.}(2017)\citenamefont {Kaneta},
  \citenamefont {Kang},\ and\ \citenamefont {Lee}}]{Kaneta:2016vkq}%
  \BibitemOpen
  \bibfield  {author} {\bibinfo {author} {\bibfnamefont {K.}~\bibnamefont
  {Kaneta}}, \bibinfo {author} {\bibfnamefont {Z.}~\bibnamefont {Kang}}, \ and\
  \bibinfo {author} {\bibfnamefont {H.-S.}\ \bibnamefont {Lee}},\ }\bibfield
  {title} {\enquote {\bibinfo {title} {{Right-handed neutrino dark matter under
  the $B − L$ gauge interaction}},}\ }\href {\doibase
  10.1007/JHEP02(2017)031} {\bibfield  {journal} {\bibinfo  {journal} {JHEP}\
  }\textbf {\bibinfo {volume} {02}},\ \bibinfo {pages} {031} (\bibinfo {year}
  {2017})},\ \Eprint {http://arxiv.org/abs/1606.09317} {arXiv:1606.09317
  [hep-ph]} \BibitemShut {NoStop}%
\bibitem [{\citenamefont {De~Romeri}\ \emph {et~al.}(2017)\citenamefont
  {De~Romeri}, \citenamefont {Fernandez-Martinez}, \citenamefont {Gehrlein},
  \citenamefont {Machado},\ and\ \citenamefont {Niro}}]{DeRomeri:2017oxa}%
  \BibitemOpen
  \bibfield  {author} {\bibinfo {author} {\bibfnamefont {V.}~\bibnamefont
  {De~Romeri}}, \bibinfo {author} {\bibfnamefont {E.}~\bibnamefont
  {Fernandez-Martinez}}, \bibinfo {author} {\bibfnamefont {J.}~\bibnamefont
  {Gehrlein}}, \bibinfo {author} {\bibfnamefont {P.~A.~N.}\ \bibnamefont
  {Machado}}, \ and\ \bibinfo {author} {\bibfnamefont {V.}~\bibnamefont
  {Niro}},\ }\bibfield  {title} {\enquote {\bibinfo {title} {{Dark Matter and
  the elusive $Z^\prime$ in a dynamical Inverse Seesaw scenario}},}\ }\href
  {\doibase 10.1007/JHEP10(2017)169} {\bibfield  {journal} {\bibinfo  {journal}
  {JHEP}\ }\textbf {\bibinfo {volume} {10}},\ \bibinfo {pages} {169} (\bibinfo
  {year} {2017})},\ \Eprint {http://arxiv.org/abs/1707.08606} {arXiv:1707.08606
  [hep-ph]} \BibitemShut {NoStop}%
\bibitem [{\citenamefont {Okada}(2018)}]{Okada:2018ktp}%
  \BibitemOpen
  \bibfield  {author} {\bibinfo {author} {\bibfnamefont {S.}~\bibnamefont
  {Okada}},\ }\bibfield  {title} {\enquote {\bibinfo {title} {{Z' Portal Dark
  Matter in the Minimal $B-L$ Model}},}\ }\href {\doibase 10.1155/2018/5340935}
  {\bibfield  {journal} {\bibinfo  {journal} {Adv. High Energy Phys.}\ }\textbf
  {\bibinfo {volume} {2018}},\ \bibinfo {pages} {5340935} (\bibinfo {year}
  {2018})},\ \Eprint {http://arxiv.org/abs/1803.06793} {arXiv:1803.06793
  [hep-ph]} \BibitemShut {NoStop}%
\bibitem [{\citenamefont {Escudero}\ \emph {et~al.}(2018)\citenamefont
  {Escudero}, \citenamefont {Witte},\ and\ \citenamefont
  {Rius}}]{Escudero:2018fwn}%
  \BibitemOpen
  \bibfield  {author} {\bibinfo {author} {\bibfnamefont {M.}~\bibnamefont
  {Escudero}}, \bibinfo {author} {\bibfnamefont {S.~J.}\ \bibnamefont {Witte}},
  \ and\ \bibinfo {author} {\bibfnamefont {N.}~\bibnamefont {Rius}},\
  }\bibfield  {title} {\enquote {\bibinfo {title} {{The dispirited case of
  gauged U(1)$_{B-L}$ dark matter}},}\ }\href {\doibase
  10.1007/JHEP08(2018)190} {\bibfield  {journal} {\bibinfo  {journal} {JHEP}\
  }\textbf {\bibinfo {volume} {08}},\ \bibinfo {pages} {190} (\bibinfo {year}
  {2018})},\ \Eprint {http://arxiv.org/abs/1806.02823} {arXiv:1806.02823
  [hep-ph]} \BibitemShut {NoStop}%
\bibitem [{\citenamefont {Ma}(2006)}]{Ma:2006km}%
  \BibitemOpen
  \bibfield  {author} {\bibinfo {author} {\bibfnamefont {E.}~\bibnamefont
  {Ma}},\ }\bibfield  {title} {\enquote {\bibinfo {title} {{Verifiable
  radiative seesaw mechanism of neutrino mass and dark matter}},}\ }\href
  {\doibase 10.1103/PhysRevD.73.077301} {\bibfield  {journal} {\bibinfo
  {journal} {Phys. Rev.}\ }\textbf {\bibinfo {volume} {D73}},\ \bibinfo {pages}
  {077301} (\bibinfo {year} {2006})},\ \Eprint
  {http://arxiv.org/abs/hep-ph/0601225} {arXiv:hep-ph/0601225 [hep-ph]}
  \BibitemShut {NoStop}%
\bibitem [{\citenamefont {Boehm}\ \emph {et~al.}(2008)\citenamefont {Boehm},
  \citenamefont {Farzan}, \citenamefont {Hambye}, \citenamefont
  {Palomares-Ruiz},\ and\ \citenamefont {Pascoli}}]{Boehm:2006mi}%
  \BibitemOpen
  \bibfield  {author} {\bibinfo {author} {\bibfnamefont {C.}~\bibnamefont
  {Boehm}}, \bibinfo {author} {\bibfnamefont {Y.}~\bibnamefont {Farzan}},
  \bibinfo {author} {\bibfnamefont {T.}~\bibnamefont {Hambye}}, \bibinfo
  {author} {\bibfnamefont {S.}~\bibnamefont {Palomares-Ruiz}}, \ and\ \bibinfo
  {author} {\bibfnamefont {S.}~\bibnamefont {Pascoli}},\ }\bibfield  {title}
  {\enquote {\bibinfo {title} {{Is it possible to explain neutrino masses with
  scalar dark matter?}}}\ }\href {\doibase 10.1103/PhysRevD.77.043516}
  {\bibfield  {journal} {\bibinfo  {journal} {Phys. Rev.}\ }\textbf {\bibinfo
  {volume} {D77}},\ \bibinfo {pages} {043516} (\bibinfo {year} {2008})},\
  \Eprint {http://arxiv.org/abs/hep-ph/0612228} {arXiv:hep-ph/0612228 [hep-ph]}
  \BibitemShut {NoStop}%
\bibitem [{\citenamefont {Farzan}\ \emph {et~al.}(2010)\citenamefont {Farzan},
  \citenamefont {Pascoli},\ and\ \citenamefont {Schmidt}}]{Farzan:2010mr}%
  \BibitemOpen
  \bibfield  {author} {\bibinfo {author} {\bibfnamefont {Y.}~\bibnamefont
  {Farzan}}, \bibinfo {author} {\bibfnamefont {S.}~\bibnamefont {Pascoli}}, \
  and\ \bibinfo {author} {\bibfnamefont {M.~A.}\ \bibnamefont {Schmidt}},\
  }\bibfield  {title} {\enquote {\bibinfo {title} {{AMEND: A model explaining
  neutrino masses and dark matter testable at the LHC and MEG}},}\ }\href
  {\doibase 10.1007/JHEP10(2010)111} {\bibfield  {journal} {\bibinfo  {journal}
  {JHEP}\ }\textbf {\bibinfo {volume} {10}},\ \bibinfo {pages} {111} (\bibinfo
  {year} {2010})},\ \Eprint {http://arxiv.org/abs/1005.5323} {arXiv:1005.5323
  [hep-ph]} \BibitemShut {NoStop}%
\bibitem [{\citenamefont {Batell}\ \emph
  {et~al.}(2018{\natexlab{a}})\citenamefont {Batell}, \citenamefont {Han},\
  and\ \citenamefont {Shams Es~Haghi}}]{Batell:2017rol}%
  \BibitemOpen
  \bibfield  {author} {\bibinfo {author} {\bibfnamefont {B.}~\bibnamefont
  {Batell}}, \bibinfo {author} {\bibfnamefont {T.}~\bibnamefont {Han}}, \ and\
  \bibinfo {author} {\bibfnamefont {B.}~\bibnamefont {Shams Es~Haghi}},\
  }\bibfield  {title} {\enquote {\bibinfo {title} {{Indirect Detection of
  Neutrino Portal Dark Matter}},}\ }\href {\doibase 10.1103/PhysRevD.97.095020}
  {\bibfield  {journal} {\bibinfo  {journal} {Phys. Rev.}\ }\textbf {\bibinfo
  {volume} {D97}},\ \bibinfo {pages} {095020} (\bibinfo {year}
  {2018}{\natexlab{a}})},\ \Eprint {http://arxiv.org/abs/1704.08708}
  {arXiv:1704.08708 [hep-ph]} \BibitemShut {NoStop}%
\bibitem [{\citenamefont {Batell}\ \emph
  {et~al.}(2018{\natexlab{b}})\citenamefont {Batell}, \citenamefont {Han},
  \citenamefont {McKeen},\ and\ \citenamefont {Shams
  Es~Haghi}}]{Batell:2017cmf}%
  \BibitemOpen
  \bibfield  {author} {\bibinfo {author} {\bibfnamefont {B.}~\bibnamefont
  {Batell}}, \bibinfo {author} {\bibfnamefont {T.}~\bibnamefont {Han}},
  \bibinfo {author} {\bibfnamefont {D.}~\bibnamefont {McKeen}}, \ and\ \bibinfo
  {author} {\bibfnamefont {B.}~\bibnamefont {Shams Es~Haghi}},\ }\bibfield
  {title} {\enquote {\bibinfo {title} {{Thermal Dark Matter Through the Dirac
  Neutrino Portal}},}\ }\href {\doibase 10.1103/PhysRevD.97.075016} {\bibfield
  {journal} {\bibinfo  {journal} {Phys. Rev.}\ }\textbf {\bibinfo {volume}
  {D97}},\ \bibinfo {pages} {075016} (\bibinfo {year} {2018}{\natexlab{b}})},\
  \Eprint {http://arxiv.org/abs/1709.07001} {arXiv:1709.07001 [hep-ph]}
  \BibitemShut {NoStop}%
\bibitem [{\citenamefont {Blennow}\ \emph {et~al.}(2019)\citenamefont
  {Blennow}, \citenamefont {Fernandez-Martinez}, \citenamefont
  {Olivares-Del~Campo}, \citenamefont {Pascoli}, \citenamefont
  {Rosauro-Alcaraz},\ and\ \citenamefont {Titov}}]{Blennow:2019fhy}%
  \BibitemOpen
  \bibfield  {author} {\bibinfo {author} {\bibfnamefont {M.}~\bibnamefont
  {Blennow}}, \bibinfo {author} {\bibfnamefont {E.}~\bibnamefont
  {Fernandez-Martinez}}, \bibinfo {author} {\bibfnamefont {A.}~\bibnamefont
  {Olivares-Del~Campo}}, \bibinfo {author} {\bibfnamefont {S.}~\bibnamefont
  {Pascoli}}, \bibinfo {author} {\bibfnamefont {S.}~\bibnamefont
  {Rosauro-Alcaraz}}, \ and\ \bibinfo {author} {\bibfnamefont {A.~V.}\
  \bibnamefont {Titov}},\ }\bibfield  {title} {\enquote {\bibinfo {title}
  {{Neutrino Portals to Dark Matter}},}\ }\href {\doibase
  10.1140/epjc/s10052-019-7060-5} {\bibfield  {journal} {\bibinfo  {journal}
  {Eur. Phys. J.}\ }\textbf {\bibinfo {volume} {C79}},\ \bibinfo {pages} {555}
  (\bibinfo {year} {2019})},\ \Eprint {http://arxiv.org/abs/1903.00006}
  {arXiv:1903.00006 [hep-ph]} \BibitemShut {NoStop}%
\bibitem [{\citenamefont {Fileviez~Perez}\ and\ \citenamefont
  {Wise}(2011)}]{FileviezPerez:2011pt}%
  \BibitemOpen
  \bibfield  {author} {\bibinfo {author} {\bibfnamefont {P.}~\bibnamefont
  {Fileviez~Perez}}\ and\ \bibinfo {author} {\bibfnamefont {M.~B.}\
  \bibnamefont {Wise}},\ }\bibfield  {title} {\enquote {\bibinfo {title}
  {{Breaking Local Baryon and Lepton Number at the TeV Scale}},}\ }\href
  {\doibase 10.1007/JHEP08(2011)068} {\bibfield  {journal} {\bibinfo  {journal}
  {JHEP}\ }\textbf {\bibinfo {volume} {08}},\ \bibinfo {pages} {068} (\bibinfo
  {year} {2011})},\ \Eprint {http://arxiv.org/abs/1106.0343} {arXiv:1106.0343
  [hep-ph]} \BibitemShut {NoStop}%
\bibitem [{\citenamefont {Duerr}\ \emph {et~al.}(2013)\citenamefont {Duerr},
  \citenamefont {Fileviez~Perez},\ and\ \citenamefont {Wise}}]{Duerr:2013dza}%
  \BibitemOpen
  \bibfield  {author} {\bibinfo {author} {\bibfnamefont {M.}~\bibnamefont
  {Duerr}}, \bibinfo {author} {\bibfnamefont {P.}~\bibnamefont
  {Fileviez~Perez}}, \ and\ \bibinfo {author} {\bibfnamefont {M.~B.}\
  \bibnamefont {Wise}},\ }\bibfield  {title} {\enquote {\bibinfo {title}
  {{Gauge Theory for Baryon and Lepton Numbers with Leptoquarks}},}\ }\href
  {\doibase 10.1103/PhysRevLett.110.231801} {\bibfield  {journal} {\bibinfo
  {journal} {Phys. Rev. Lett.}\ }\textbf {\bibinfo {volume} {110}},\ \bibinfo
  {pages} {231801} (\bibinfo {year} {2013})},\ \Eprint
  {http://arxiv.org/abs/1304.0576} {arXiv:1304.0576 [hep-ph]} \BibitemShut
  {NoStop}%
\bibitem [{\citenamefont {Schwaller}\ \emph {et~al.}(2013)\citenamefont
  {Schwaller}, \citenamefont {Tait},\ and\ \citenamefont
  {Vega-Morales}}]{Schwaller:2013hqa}%
  \BibitemOpen
  \bibfield  {author} {\bibinfo {author} {\bibfnamefont {P.}~\bibnamefont
  {Schwaller}}, \bibinfo {author} {\bibfnamefont {T.~M.~P.}\ \bibnamefont
  {Tait}}, \ and\ \bibinfo {author} {\bibfnamefont {R.}~\bibnamefont
  {Vega-Morales}},\ }\bibfield  {title} {\enquote {\bibinfo {title} {{Dark
  Matter and Vectorlike Leptons from Gauged Lepton Number}},}\ }\href {\doibase
  10.1103/PhysRevD.88.035001} {\bibfield  {journal} {\bibinfo  {journal} {Phys.
  Rev.}\ }\textbf {\bibinfo {volume} {D88}},\ \bibinfo {pages} {035001}
  (\bibinfo {year} {2013})},\ \Eprint {http://arxiv.org/abs/1305.1108}
  {arXiv:1305.1108 [hep-ph]} \BibitemShut {NoStop}%
\bibitem [{\citenamefont {Fileviez~Perez}\ \emph {et~al.}(2014)\citenamefont
  {Fileviez~Perez}, \citenamefont {Ohmer},\ and\ \citenamefont
  {Patel}}]{Perez:2014qfa}%
  \BibitemOpen
  \bibfield  {author} {\bibinfo {author} {\bibfnamefont {P.}~\bibnamefont
  {Fileviez~Perez}}, \bibinfo {author} {\bibfnamefont {S.}~\bibnamefont
  {Ohmer}}, \ and\ \bibinfo {author} {\bibfnamefont {H.~H.}\ \bibnamefont
  {Patel}},\ }\bibfield  {title} {\enquote {\bibinfo {title} {{Minimal Theory
  for Lepto-Baryons}},}\ }\href {\doibase 10.1016/j.physletb.2014.06.057}
  {\bibfield  {journal} {\bibinfo  {journal} {Phys. Lett.}\ }\textbf {\bibinfo
  {volume} {B735}},\ \bibinfo {pages} {283--287} (\bibinfo {year} {2014})},\
  \Eprint {http://arxiv.org/abs/1403.8029} {arXiv:1403.8029 [hep-ph]}
  \BibitemShut {NoStop}%
\bibitem [{\citenamefont {Chao}(2011)}]{Chao:2010mp}%
  \BibitemOpen
  \bibfield  {author} {\bibinfo {author} {\bibfnamefont {W.}~\bibnamefont
  {Chao}},\ }\bibfield  {title} {\enquote {\bibinfo {title} {{Pure Leptonic
  Gauge Symmetry, Neutrino Masses and Dark Matter}},}\ }\href {\doibase
  10.1016/j.physletb.2010.10.056} {\bibfield  {journal} {\bibinfo  {journal}
  {Phys. Lett.}\ }\textbf {\bibinfo {volume} {B695}},\ \bibinfo {pages}
  {157--161} (\bibinfo {year} {2011})},\ \Eprint
  {http://arxiv.org/abs/1005.1024} {arXiv:1005.1024 [hep-ph]} \BibitemShut
  {NoStop}%
\bibitem [{\citenamefont {Aranda}\ \emph {et~al.}(2015)\citenamefont {Aranda},
  \citenamefont {Jiménez},\ and\ \citenamefont
  {Vaquera-Araujo}}]{Aranda:2014zta}%
  \BibitemOpen
  \bibfield  {author} {\bibinfo {author} {\bibfnamefont {A.}~\bibnamefont
  {Aranda}}, \bibinfo {author} {\bibfnamefont {E.}~\bibnamefont {Jiménez}}, \
  and\ \bibinfo {author} {\bibfnamefont {C.~A.}\ \bibnamefont
  {Vaquera-Araujo}},\ }\bibfield  {title} {\enquote {\bibinfo {title}
  {{Electroweak phase transition in a model with gauged lepton number}},}\
  }\href {\doibase 10.1007/JHEP01(2015)070} {\bibfield  {journal} {\bibinfo
  {journal} {JHEP}\ }\textbf {\bibinfo {volume} {01}},\ \bibinfo {pages} {070}
  (\bibinfo {year} {2015})},\ \Eprint {http://arxiv.org/abs/1410.7508}
  {arXiv:1410.7508 [hep-ph]} \BibitemShut {NoStop}%
\bibitem [{\citenamefont {Fornal}\ \emph {et~al.}(2017)\citenamefont {Fornal},
  \citenamefont {Shirman}, \citenamefont {Tait},\ and\ \citenamefont
  {West}}]{Fornal:2017owa}%
  \BibitemOpen
  \bibfield  {author} {\bibinfo {author} {\bibfnamefont {B.}~\bibnamefont
  {Fornal}}, \bibinfo {author} {\bibfnamefont {Y.}~\bibnamefont {Shirman}},
  \bibinfo {author} {\bibfnamefont {T.~M.~P.}\ \bibnamefont {Tait}}, \ and\
  \bibinfo {author} {\bibfnamefont {J.~R.}\ \bibnamefont {West}},\ }\bibfield
  {title} {\enquote {\bibinfo {title} {{Asymmetric dark matter and baryogenesis
  from $SU(2)_{\ell}$}},}\ }\href {\doibase 10.1103/PhysRevD.96.035001}
  {\bibfield  {journal} {\bibinfo  {journal} {Phys. Rev.}\ }\textbf {\bibinfo
  {volume} {D96}},\ \bibinfo {pages} {035001} (\bibinfo {year} {2017})},\
  \Eprint {http://arxiv.org/abs/1703.00199} {arXiv:1703.00199 [hep-ph]}
  \BibitemShut {NoStop}%
\bibitem [{\citenamefont {Chang}\ and\ \citenamefont
  {Ng}(2018)}]{Chang:2018vdd}%
  \BibitemOpen
  \bibfield  {author} {\bibinfo {author} {\bibfnamefont {W.-F.}\ \bibnamefont
  {Chang}}\ and\ \bibinfo {author} {\bibfnamefont {J.~N.}\ \bibnamefont {Ng}},\
  }\bibfield  {title} {\enquote {\bibinfo {title} {{Study of Gauged Lepton
  Symmetry Signatures at Colliders}},}\ }\href {\doibase
  10.1103/PhysRevD.98.035015} {\bibfield  {journal} {\bibinfo  {journal} {Phys.
  Rev.}\ }\textbf {\bibinfo {volume} {D98}},\ \bibinfo {pages} {035015}
  (\bibinfo {year} {2018})},\ \Eprint {http://arxiv.org/abs/1805.10382}
  {arXiv:1805.10382 [hep-ph]} \BibitemShut {NoStop}%
\bibitem [{\citenamefont {Chang}\ and\ \citenamefont
  {Ng}(2019)}]{Chang:2018nid}%
  \BibitemOpen
  \bibfield  {author} {\bibinfo {author} {\bibfnamefont {W.-F.}\ \bibnamefont
  {Chang}}\ and\ \bibinfo {author} {\bibfnamefont {J.~N.}\ \bibnamefont {Ng}},\
  }\bibfield  {title} {\enquote {\bibinfo {title} {{Alternative Perspective on
  Gauged Lepton Number and Implications for Collider Physics}},}\ }\href
  {\doibase 10.1103/PhysRevD.99.075025} {\bibfield  {journal} {\bibinfo
  {journal} {Phys. Rev.}\ }\textbf {\bibinfo {volume} {D99}},\ \bibinfo {pages}
  {075025} (\bibinfo {year} {2019})},\ \Eprint
  {http://arxiv.org/abs/1808.08188} {arXiv:1808.08188 [hep-ph]} \BibitemShut
  {NoStop}%
\bibitem [{\citenamefont {Alioli}\ \emph {et~al.}(2018)\citenamefont {Alioli},
  \citenamefont {Farina}, \citenamefont {Pappadopulo},\ and\ \citenamefont
  {Ruderman}}]{Alioli:2017nzr}%
  \BibitemOpen
  \bibfield  {author} {\bibinfo {author} {\bibfnamefont {S.}~\bibnamefont
  {Alioli}}, \bibinfo {author} {\bibfnamefont {M.}~\bibnamefont {Farina}},
  \bibinfo {author} {\bibfnamefont {D.}~\bibnamefont {Pappadopulo}}, \ and\
  \bibinfo {author} {\bibfnamefont {J.~T.}\ \bibnamefont {Ruderman}},\
  }\bibfield  {title} {\enquote {\bibinfo {title} {{Catching a New Force by the
  Tail}},}\ }\href {\doibase 10.1103/PhysRevLett.120.101801} {\bibfield
  {journal} {\bibinfo  {journal} {Phys. Rev. Lett.}\ }\textbf {\bibinfo
  {volume} {120}},\ \bibinfo {pages} {101801} (\bibinfo {year} {2018})},\
  \Eprint {http://arxiv.org/abs/1712.02347} {arXiv:1712.02347 [hep-ph]}
  \BibitemShut {NoStop}%
\bibitem [{\citenamefont {Aaboud}\ \emph {et~al.}(2017)\citenamefont {Aaboud}
  \emph {et~al.}}]{Aaboud:2017buh}%
  \BibitemOpen
  \bibfield  {author} {\bibinfo {author} {\bibfnamefont {M.}~\bibnamefont
  {Aaboud}} \emph {et~al.} (\bibinfo {collaboration} {ATLAS}),\ }\bibfield
  {title} {\enquote {\bibinfo {title} {{Search for new high-mass phenomena in
  the dilepton final state using 36 fb of proton-proton collision data at 13
  TeV with the ATLAS detector}},}\ }\href {\doibase 10.1007/JHEP10(2017)182}
  {\bibfield  {journal} {\bibinfo  {journal} {JHEP}\ }\textbf {\bibinfo
  {volume} {10}},\ \bibinfo {pages} {182} (\bibinfo {year} {2017})},\ \Eprint
  {http://arxiv.org/abs/1707.02424} {arXiv:1707.02424 [hep-ex]} \BibitemShut
  {NoStop}%
\bibitem [{\citenamefont {Barger}\ \emph {et~al.}(2003)\citenamefont {Barger},
  \citenamefont {Langacker},\ and\ \citenamefont {Lee}}]{Barger:2003zh}%
  \BibitemOpen
  \bibfield  {author} {\bibinfo {author} {\bibfnamefont {V.}~\bibnamefont
  {Barger}}, \bibinfo {author} {\bibfnamefont {P.}~\bibnamefont {Langacker}}, \
  and\ \bibinfo {author} {\bibfnamefont {H.-S.}\ \bibnamefont {Lee}},\
  }\bibfield  {title} {\enquote {\bibinfo {title} {{Primordial nucleosynthesis
  constraints on $Z^\prime$ properties}},}\ }\href {\doibase
  10.1103/PhysRevD.67.075009} {\bibfield  {journal} {\bibinfo  {journal} {Phys.
  Rev.}\ }\textbf {\bibinfo {volume} {D67}},\ \bibinfo {pages} {075009}
  (\bibinfo {year} {2003})},\ \Eprint {http://arxiv.org/abs/hep-ph/0302066}
  {arXiv:hep-ph/0302066 [hep-ph]} \BibitemShut {NoStop}%
\bibitem [{\citenamefont {Hamann}\ \emph {et~al.}(2011)\citenamefont {Hamann},
  \citenamefont {Hannestad}, \citenamefont {Raffelt},\ and\ \citenamefont
  {Wong}}]{Hamann:2011ge}%
  \BibitemOpen
  \bibfield  {author} {\bibinfo {author} {\bibfnamefont {J.}~\bibnamefont
  {Hamann}}, \bibinfo {author} {\bibfnamefont {S.}~\bibnamefont {Hannestad}},
  \bibinfo {author} {\bibfnamefont {G.~G.}\ \bibnamefont {Raffelt}}, \ and\
  \bibinfo {author} {\bibfnamefont {Y.~Y.~Y.}\ \bibnamefont {Wong}},\
  }\bibfield  {title} {\enquote {\bibinfo {title} {{Sterile neutrinos with eV
  masses in cosmology: How disfavoured exactly?}}}\ }\href {\doibase
  10.1088/1475-7516/2011/09/034} {\bibfield  {journal} {\bibinfo  {journal}
  {JCAP}\ }\textbf {\bibinfo {volume} {1109}},\ \bibinfo {pages} {034}
  (\bibinfo {year} {2011})},\ \Eprint {http://arxiv.org/abs/1108.4136}
  {arXiv:1108.4136 [astro-ph.CO]} \BibitemShut {NoStop}%
\bibitem [{\citenamefont {Anchordoqui}\ and\ \citenamefont
  {Goldberg}(2012)}]{Anchordoqui:2011nh}%
  \BibitemOpen
  \bibfield  {author} {\bibinfo {author} {\bibfnamefont {L.~A.}\ \bibnamefont
  {Anchordoqui}}\ and\ \bibinfo {author} {\bibfnamefont {H.}~\bibnamefont
  {Goldberg}},\ }\bibfield  {title} {\enquote {\bibinfo {title} {{Neutrino
  cosmology after WMAP 7-Year data and LHC first Z' bounds}},}\ }\href
  {\doibase 10.1103/PhysRevLett.108.081805} {\bibfield  {journal} {\bibinfo
  {journal} {Phys. Rev. Lett.}\ }\textbf {\bibinfo {volume} {108}},\ \bibinfo
  {pages} {081805} (\bibinfo {year} {2012})},\ \Eprint
  {http://arxiv.org/abs/1111.7264} {arXiv:1111.7264 [hep-ph]} \BibitemShut
  {NoStop}%
\bibitem [{\citenamefont {Hannestad}\ \emph {et~al.}(2012)\citenamefont
  {Hannestad}, \citenamefont {Tamborra},\ and\ \citenamefont
  {Tram}}]{Hannestad:2012ky}%
  \BibitemOpen
  \bibfield  {author} {\bibinfo {author} {\bibfnamefont {S.}~\bibnamefont
  {Hannestad}}, \bibinfo {author} {\bibfnamefont {I.}~\bibnamefont {Tamborra}},
  \ and\ \bibinfo {author} {\bibfnamefont {T.}~\bibnamefont {Tram}},\
  }\bibfield  {title} {\enquote {\bibinfo {title} {{Thermalisation of light
  sterile neutrinos in the early universe}},}\ }\href {\doibase
  10.1088/1475-7516/2012/07/025} {\bibfield  {journal} {\bibinfo  {journal}
  {JCAP}\ }\textbf {\bibinfo {volume} {1207}},\ \bibinfo {pages} {025}
  (\bibinfo {year} {2012})},\ \Eprint {http://arxiv.org/abs/1204.5861}
  {arXiv:1204.5861 [astro-ph.CO]} \BibitemShut {NoStop}%
\bibitem [{\citenamefont {Solaguren-Beascoa}\ and\ \citenamefont
  {Gonzalez-Garcia}(2013)}]{SolagurenBeascoa:2012cz}%
  \BibitemOpen
  \bibfield  {author} {\bibinfo {author} {\bibfnamefont {A.}~\bibnamefont
  {Solaguren-Beascoa}}\ and\ \bibinfo {author} {\bibfnamefont {M.~C.}\
  \bibnamefont {Gonzalez-Garcia}},\ }\bibfield  {title} {\enquote {\bibinfo
  {title} {{Dark Radiation Confronting LHC in Z' Models}},}\ }\href {\doibase
  10.1016/j.physletb.2012.12.065} {\bibfield  {journal} {\bibinfo  {journal}
  {Phys. Lett.}\ }\textbf {\bibinfo {volume} {B719}},\ \bibinfo {pages}
  {121--125} (\bibinfo {year} {2013})},\ \Eprint
  {http://arxiv.org/abs/1210.6350} {arXiv:1210.6350 [hep-ph]} \BibitemShut
  {NoStop}%
\bibitem [{\citenamefont {Anchordoqui}\ \emph {et~al.}(2013)\citenamefont
  {Anchordoqui}, \citenamefont {Goldberg},\ and\ \citenamefont
  {Steigman}}]{Anchordoqui:2012qu}%
  \BibitemOpen
  \bibfield  {author} {\bibinfo {author} {\bibfnamefont {L.~A.}\ \bibnamefont
  {Anchordoqui}}, \bibinfo {author} {\bibfnamefont {H.}~\bibnamefont
  {Goldberg}}, \ and\ \bibinfo {author} {\bibfnamefont {G.}~\bibnamefont
  {Steigman}},\ }\bibfield  {title} {\enquote {\bibinfo {title} {{Right-Handed
  Neutrinos as the Dark Radiation: Status and Forecasts for the LHC}},}\ }\href
  {\doibase 10.1016/j.physletb.2012.12.019} {\bibfield  {journal} {\bibinfo
  {journal} {Phys. Lett.}\ }\textbf {\bibinfo {volume} {B718}},\ \bibinfo
  {pages} {1162--1165} (\bibinfo {year} {2013})},\ \Eprint
  {http://arxiv.org/abs/1211.0186} {arXiv:1211.0186 [hep-ph]} \BibitemShut
  {NoStop}%
\bibitem [{\citenamefont {Ho}\ and\ \citenamefont
  {Scherrer}(2013)}]{Ho:2012ug}%
  \BibitemOpen
  \bibfield  {author} {\bibinfo {author} {\bibfnamefont {C.~M.}\ \bibnamefont
  {Ho}}\ and\ \bibinfo {author} {\bibfnamefont {R.~J.}\ \bibnamefont
  {Scherrer}},\ }\bibfield  {title} {\enquote {\bibinfo {title} {{Limits on MeV
  Dark Matter from the Effective Number of Neutrinos}},}\ }\href {\doibase
  10.1103/PhysRevD.87.023505} {\bibfield  {journal} {\bibinfo  {journal} {Phys.
  Rev.}\ }\textbf {\bibinfo {volume} {D87}},\ \bibinfo {pages} {023505}
  (\bibinfo {year} {2013})},\ \Eprint {http://arxiv.org/abs/1208.4347}
  {arXiv:1208.4347 [astro-ph.CO]} \BibitemShut {NoStop}%
\bibitem [{\citenamefont {Brust}\ \emph {et~al.}(2013)\citenamefont {Brust},
  \citenamefont {Kaplan},\ and\ \citenamefont {Walters}}]{Brust:2013xpv}%
  \BibitemOpen
  \bibfield  {author} {\bibinfo {author} {\bibfnamefont {C.}~\bibnamefont
  {Brust}}, \bibinfo {author} {\bibfnamefont {D.~E.}\ \bibnamefont {Kaplan}}, \
  and\ \bibinfo {author} {\bibfnamefont {M.~T.}\ \bibnamefont {Walters}},\
  }\bibfield  {title} {\enquote {\bibinfo {title} {{New Light Species and the
  CMB}},}\ }\href {\doibase 10.1007/JHEP12(2013)058} {\bibfield  {journal}
  {\bibinfo  {journal} {JHEP}\ }\textbf {\bibinfo {volume} {12}},\ \bibinfo
  {pages} {058} (\bibinfo {year} {2013})},\ \Eprint
  {http://arxiv.org/abs/1303.5379} {arXiv:1303.5379 [hep-ph]} \BibitemShut
  {NoStop}%
\bibitem [{\citenamefont {Boehm}\ \emph {et~al.}(2013)\citenamefont {Boehm},
  \citenamefont {Dolan},\ and\ \citenamefont {McCabe}}]{Boehm:2013jpa}%
  \BibitemOpen
  \bibfield  {author} {\bibinfo {author} {\bibfnamefont {C.}~\bibnamefont
  {Boehm}}, \bibinfo {author} {\bibfnamefont {M.~J.}\ \bibnamefont {Dolan}}, \
  and\ \bibinfo {author} {\bibfnamefont {C.}~\bibnamefont {McCabe}},\
  }\bibfield  {title} {\enquote {\bibinfo {title} {{A Lower Bound on the Mass
  of Cold Thermal Dark Matter from Planck}},}\ }\href {\doibase
  10.1088/1475-7516/2013/08/041} {\bibfield  {journal} {\bibinfo  {journal}
  {JCAP}\ }\textbf {\bibinfo {volume} {1308}},\ \bibinfo {pages} {041}
  (\bibinfo {year} {2013})},\ \Eprint {http://arxiv.org/abs/1303.6270}
  {arXiv:1303.6270 [hep-ph]} \BibitemShut {NoStop}%
\bibitem [{\citenamefont {Fileviez~Perez}\ and\ \citenamefont
  {Spinner}(2014)}]{Perez:2013kla}%
  \BibitemOpen
  \bibfield  {author} {\bibinfo {author} {\bibfnamefont {P.}~\bibnamefont
  {Fileviez~Perez}}\ and\ \bibinfo {author} {\bibfnamefont {S.}~\bibnamefont
  {Spinner}},\ }\bibfield  {title} {\enquote {\bibinfo {title} {{Supersymmetry
  at the LHC and The Theory of R-parity}},}\ }\href {\doibase
  10.1016/j.physletb.2013.12.022} {\bibfield  {journal} {\bibinfo  {journal}
  {Phys. Lett.}\ }\textbf {\bibinfo {volume} {B728}},\ \bibinfo {pages}
  {489--495} (\bibinfo {year} {2014})},\ \Eprint
  {http://arxiv.org/abs/1308.0524} {arXiv:1308.0524 [hep-ph]} \BibitemShut
  {NoStop}%
\bibitem [{\citenamefont {Escudero}(2019)}]{Escudero:2018mvt}%
  \BibitemOpen
  \bibfield  {author} {\bibinfo {author} {\bibfnamefont {M.}~\bibnamefont
  {Escudero}},\ }\bibfield  {title} {\enquote {\bibinfo {title} {{Neutrino
  decoupling beyond the Standard Model: CMB constraints on the Dark Matter mass
  with a fast and precise $N_{\rm eff}$ evaluation}},}\ }\href {\doibase
  10.1088/1475-7516/2019/02/007} {\bibfield  {journal} {\bibinfo  {journal}
  {JCAP}\ }\textbf {\bibinfo {volume} {1902}},\ \bibinfo {pages} {007}
  (\bibinfo {year} {2019})},\ \Eprint {http://arxiv.org/abs/1812.05605}
  {arXiv:1812.05605 [hep-ph]} \BibitemShut {NoStop}%
\bibitem [{\citenamefont {Cadamuro}\ \emph {et~al.}(2011)\citenamefont
  {Cadamuro}, \citenamefont {Hannestad}, \citenamefont {Raffelt},\ and\
  \citenamefont {Redondo}}]{Cadamuro:2010cz}%
  \BibitemOpen
  \bibfield  {author} {\bibinfo {author} {\bibfnamefont {D.}~\bibnamefont
  {Cadamuro}}, \bibinfo {author} {\bibfnamefont {S.}~\bibnamefont {Hannestad}},
  \bibinfo {author} {\bibfnamefont {G.}~\bibnamefont {Raffelt}}, \ and\
  \bibinfo {author} {\bibfnamefont {J.}~\bibnamefont {Redondo}},\ }\bibfield
  {title} {\enquote {\bibinfo {title} {{Cosmological bounds on sub-MeV mass
  axions}},}\ }\href {\doibase 10.1088/1475-7516/2011/02/003} {\bibfield
  {journal} {\bibinfo  {journal} {JCAP}\ }\textbf {\bibinfo {volume} {1102}},\
  \bibinfo {pages} {003} (\bibinfo {year} {2011})},\ \Eprint
  {http://arxiv.org/abs/1011.3694} {arXiv:1011.3694 [hep-ph]} \BibitemShut
  {NoStop}%
\bibitem [{\citenamefont {Salvio}\ \emph {et~al.}(2014)\citenamefont {Salvio},
  \citenamefont {Strumia},\ and\ \citenamefont {Xue}}]{Salvio:2013iaa}%
  \BibitemOpen
  \bibfield  {author} {\bibinfo {author} {\bibfnamefont {A.}~\bibnamefont
  {Salvio}}, \bibinfo {author} {\bibfnamefont {A.}~\bibnamefont {Strumia}}, \
  and\ \bibinfo {author} {\bibfnamefont {W.}~\bibnamefont {Xue}},\ }\bibfield
  {title} {\enquote {\bibinfo {title} {{Thermal axion production}},}\ }\href
  {\doibase 10.1088/1475-7516/2014/01/011} {\bibfield  {journal} {\bibinfo
  {journal} {JCAP}\ }\textbf {\bibinfo {volume} {1401}},\ \bibinfo {pages}
  {011} (\bibinfo {year} {2014})},\ \Eprint {http://arxiv.org/abs/1310.6982}
  {arXiv:1310.6982 [hep-ph]} \BibitemShut {NoStop}%
\bibitem [{\citenamefont {Kawasaki}\ \emph {et~al.}(2015)\citenamefont
  {Kawasaki}, \citenamefont {Yamada},\ and\ \citenamefont
  {Yanagida}}]{Kawasaki:2015ofa}%
  \BibitemOpen
  \bibfield  {author} {\bibinfo {author} {\bibfnamefont {M.}~\bibnamefont
  {Kawasaki}}, \bibinfo {author} {\bibfnamefont {M.}~\bibnamefont {Yamada}}, \
  and\ \bibinfo {author} {\bibfnamefont {T.~T.}\ \bibnamefont {Yanagida}},\
  }\bibfield  {title} {\enquote {\bibinfo {title} {{Observable dark radiation
  from a cosmologically safe QCD axion}},}\ }\href {\doibase
  10.1103/PhysRevD.91.125018} {\bibfield  {journal} {\bibinfo  {journal} {Phys.
  Rev.}\ }\textbf {\bibinfo {volume} {D91}},\ \bibinfo {pages} {125018}
  (\bibinfo {year} {2015})},\ \Eprint {http://arxiv.org/abs/1504.04126}
  {arXiv:1504.04126 [hep-ph]} \BibitemShut {NoStop}%
\bibitem [{\citenamefont {Baumann}\ \emph {et~al.}(2016)\citenamefont
  {Baumann}, \citenamefont {Green},\ and\ \citenamefont
  {Wallisch}}]{Baumann:2016wac}%
  \BibitemOpen
  \bibfield  {author} {\bibinfo {author} {\bibfnamefont {D.}~\bibnamefont
  {Baumann}}, \bibinfo {author} {\bibfnamefont {D.}~\bibnamefont {Green}}, \
  and\ \bibinfo {author} {\bibfnamefont {B.}~\bibnamefont {Wallisch}},\
  }\bibfield  {title} {\enquote {\bibinfo {title} {{New Target for Cosmic Axion
  Searches}},}\ }\href {\doibase 10.1103/PhysRevLett.117.171301} {\bibfield
  {journal} {\bibinfo  {journal} {Phys. Rev. Lett.}\ }\textbf {\bibinfo
  {volume} {117}},\ \bibinfo {pages} {171301} (\bibinfo {year} {2016})},\
  \Eprint {http://arxiv.org/abs/1604.08614} {arXiv:1604.08614 [astro-ph.CO]}
  \BibitemShut {NoStop}%
\bibitem [{\citenamefont {Aghanim}\ \emph {et~al.}(2018)\citenamefont {Aghanim}
  \emph {et~al.}}]{Aghanim:2018eyx}%
  \BibitemOpen
  \bibfield  {author} {\bibinfo {author} {\bibfnamefont {N.}~\bibnamefont
  {Aghanim}} \emph {et~al.} (\bibinfo {collaboration} {Planck}),\ }\bibfield
  {title} {\enquote {\bibinfo {title} {{Planck 2018 results. VI. Cosmological
  parameters}},}\ }\href@noop {} {\  (\bibinfo {year} {2018})},\ \Eprint
  {http://arxiv.org/abs/1807.06209} {arXiv:1807.06209 [astro-ph.CO]}
  \BibitemShut {NoStop}%
\bibitem [{\citenamefont {Enqvist}\ \emph {et~al.}(1992)\citenamefont
  {Enqvist}, \citenamefont {Kainulainen},\ and\ \citenamefont
  {Semikoz}}]{Enqvist:1991gx}%
  \BibitemOpen
  \bibfield  {author} {\bibinfo {author} {\bibfnamefont {K.}~\bibnamefont
  {Enqvist}}, \bibinfo {author} {\bibfnamefont {K.}~\bibnamefont
  {Kainulainen}}, \ and\ \bibinfo {author} {\bibfnamefont {V.}~\bibnamefont
  {Semikoz}},\ }\bibfield  {title} {\enquote {\bibinfo {title} {{Neutrino
  annihilation in hot plasma}},}\ }\href {\doibase
  10.1016/0550-3213(92)90359-J} {\bibfield  {journal} {\bibinfo  {journal}
  {Nucl. Phys.}\ }\textbf {\bibinfo {volume} {B374}},\ \bibinfo {pages}
  {392--404} (\bibinfo {year} {1992})}\BibitemShut {NoStop}%
\bibitem [{\citenamefont {de~Salas}\ and\ \citenamefont
  {Pastor}(2016)}]{deSalas:2016ztq}%
  \BibitemOpen
  \bibfield  {author} {\bibinfo {author} {\bibfnamefont {P.~F.}\ \bibnamefont
  {de~Salas}}\ and\ \bibinfo {author} {\bibfnamefont {S.}~\bibnamefont
  {Pastor}},\ }\bibfield  {title} {\enquote {\bibinfo {title} {{Relic neutrino
  decoupling with flavour oscillations revisited}},}\ }\href {\doibase
  10.1088/1475-7516/2016/07/051} {\bibfield  {journal} {\bibinfo  {journal}
  {JCAP}\ }\textbf {\bibinfo {volume} {1607}},\ \bibinfo {pages} {051}
  (\bibinfo {year} {2016})},\ \Eprint {http://arxiv.org/abs/1606.06986}
  {arXiv:1606.06986 [hep-ph]} \BibitemShut {NoStop}%
\bibitem [{\citenamefont {Borsanyi}\ \emph {et~al.}(2016)\citenamefont
  {Borsanyi} \emph {et~al.}}]{Borsanyi:2016ksw}%
  \BibitemOpen
  \bibfield  {author} {\bibinfo {author} {\bibfnamefont {S.}~\bibnamefont
  {Borsanyi}} \emph {et~al.},\ }\bibfield  {title} {\enquote {\bibinfo {title}
  {{Calculation of the axion mass based on high-temperature lattice quantum
  chromodynamics}},}\ }\href {\doibase 10.1038/nature20115} {\bibfield
  {journal} {\bibinfo  {journal} {Nature}\ }\textbf {\bibinfo {volume} {539}},\
  \bibinfo {pages} {69--71} (\bibinfo {year} {2016})},\ \Eprint
  {http://arxiv.org/abs/1606.07494} {arXiv:1606.07494 [hep-lat]} \BibitemShut
  {NoStop}%
\bibitem [{\citenamefont {Abazajian}\ \emph {et~al.}(2016)\citenamefont
  {Abazajian} \emph {et~al.}}]{Abazajian:2016yjj}%
  \BibitemOpen
  \bibfield  {author} {\bibinfo {author} {\bibfnamefont {K.~N.}\ \bibnamefont
  {Abazajian}} \emph {et~al.} (\bibinfo {collaboration} {CMB-S4}),\ }\bibfield
  {title} {\enquote {\bibinfo {title} {{CMB-S4 Science Book, First Edition}},}\
  }\href@noop {} {\  (\bibinfo {year} {2016})},\ \Eprint
  {http://arxiv.org/abs/1610.02743} {arXiv:1610.02743 [astro-ph.CO]}
  \BibitemShut {NoStop}%
\bibitem [{\citenamefont {Ruegg}\ and\ \citenamefont
  {Ruiz-Altaba}(2004)}]{Ruegg:2003ps}%
  \BibitemOpen
  \bibfield  {author} {\bibinfo {author} {\bibfnamefont {H.}~\bibnamefont
  {Ruegg}}\ and\ \bibinfo {author} {\bibfnamefont {M.}~\bibnamefont
  {Ruiz-Altaba}},\ }\bibfield  {title} {\enquote {\bibinfo {title} {{The
  Stueckelberg field}},}\ }\href {\doibase 10.1142/S0217751X04019755}
  {\bibfield  {journal} {\bibinfo  {journal} {Int. J. Mod. Phys.}\ }\textbf
  {\bibinfo {volume} {A19}},\ \bibinfo {pages} {3265--3348} (\bibinfo {year}
  {2004})},\ \Eprint {http://arxiv.org/abs/hep-th/0304245}
  {arXiv:hep-th/0304245 [hep-th]} \BibitemShut {NoStop}%
\bibitem [{\citenamefont {Feldman}\ \emph {et~al.}(2012)\citenamefont
  {Feldman}, \citenamefont {Fileviez~Perez},\ and\ \citenamefont
  {Nath}}]{Feldman:2011ms}%
  \BibitemOpen
  \bibfield  {author} {\bibinfo {author} {\bibfnamefont {D.}~\bibnamefont
  {Feldman}}, \bibinfo {author} {\bibfnamefont {P.}~\bibnamefont
  {Fileviez~Perez}}, \ and\ \bibinfo {author} {\bibfnamefont {P.}~\bibnamefont
  {Nath}},\ }\bibfield  {title} {\enquote {\bibinfo {title} {{R-parity
  Conservation via the Stueckelberg Mechanism: LHC and Dark Matter Signals}},}\
  }\href {\doibase 10.1007/JHEP01(2012)038} {\bibfield  {journal} {\bibinfo
  {journal} {JHEP}\ }\textbf {\bibinfo {volume} {01}},\ \bibinfo {pages} {038}
  (\bibinfo {year} {2012})},\ \Eprint {http://arxiv.org/abs/1109.2901}
  {arXiv:1109.2901 [hep-ph]} \BibitemShut {NoStop}%
\bibitem [{\citenamefont {Aprile}\ \emph {et~al.}(2017)\citenamefont {Aprile}
  \emph {et~al.}}]{Aprile:2017iyp}%
  \BibitemOpen
  \bibfield  {author} {\bibinfo {author} {\bibfnamefont {E.}~\bibnamefont
  {Aprile}} \emph {et~al.} (\bibinfo {collaboration} {XENON}),\ }\bibfield
  {title} {\enquote {\bibinfo {title} {{First Dark Matter Search Results from
  the XENON1T Experiment}},}\ }\href {\doibase 10.1103/PhysRevLett.119.181301}
  {\bibfield  {journal} {\bibinfo  {journal} {Phys. Rev. Lett.}\ }\textbf
  {\bibinfo {volume} {119}},\ \bibinfo {pages} {181301} (\bibinfo {year}
  {2017})},\ \Eprint {http://arxiv.org/abs/1705.06655} {arXiv:1705.06655
  [astro-ph.CO]} \BibitemShut {NoStop}%
\bibitem [{\citenamefont {Aprile}\ \emph {et~al.}(2016)\citenamefont {Aprile}
  \emph {et~al.}}]{Aprile:2015uzo}%
  \BibitemOpen
  \bibfield  {author} {\bibinfo {author} {\bibfnamefont {E.}~\bibnamefont
  {Aprile}} \emph {et~al.} (\bibinfo {collaboration} {XENON}),\ }\bibfield
  {title} {\enquote {\bibinfo {title} {{Physics reach of the XENON1T dark
  matter experiment}},}\ }\href {\doibase 10.1088/1475-7516/2016/04/027}
  {\bibfield  {journal} {\bibinfo  {journal} {JCAP}\ }\textbf {\bibinfo
  {volume} {1604}},\ \bibinfo {pages} {027} (\bibinfo {year} {2016})},\ \Eprint
  {http://arxiv.org/abs/1512.07501} {arXiv:1512.07501 [physics.ins-det]}
  \BibitemShut {NoStop}%
\bibitem [{\citenamefont {Bélanger}\ \emph {et~al.}(2018)\citenamefont
  {Bélanger}, \citenamefont {Boudjema}, \citenamefont {Goudelis},
  \citenamefont {Pukhov},\ and\ \citenamefont {Zaldivar}}]{Belanger:2018mqt}%
  \BibitemOpen
  \bibfield  {author} {\bibinfo {author} {\bibfnamefont {G.}~\bibnamefont
  {Bélanger}}, \bibinfo {author} {\bibfnamefont {F.}~\bibnamefont {Boudjema}},
  \bibinfo {author} {\bibfnamefont {A.}~\bibnamefont {Goudelis}}, \bibinfo
  {author} {\bibfnamefont {A.}~\bibnamefont {Pukhov}}, \ and\ \bibinfo {author}
  {\bibfnamefont {B.}~\bibnamefont {Zaldivar}},\ }\bibfield  {title} {\enquote
  {\bibinfo {title} {{micrOMEGAs5.0 : Freeze-in}},}\ }\href {\doibase
  10.1016/j.cpc.2018.04.027} {\bibfield  {journal} {\bibinfo  {journal}
  {Comput. Phys. Commun.}\ }\textbf {\bibinfo {volume} {231}},\ \bibinfo
  {pages} {173--186} (\bibinfo {year} {2018})},\ \Eprint
  {http://arxiv.org/abs/1801.03509} {arXiv:1801.03509 [hep-ph]} \BibitemShut
  {NoStop}%
\bibitem [{\citenamefont {Semenov}(2016)}]{Semenov:2014rea}%
  \BibitemOpen
  \bibfield  {author} {\bibinfo {author} {\bibfnamefont {A.}~\bibnamefont
  {Semenov}},\ }\bibfield  {title} {\enquote {\bibinfo {title} {{LanHEP — A
  package for automatic generation of Feynman rules from the Lagrangian.
  Version 3.2}},}\ }\href {\doibase 10.1016/j.cpc.2016.01.003} {\bibfield
  {journal} {\bibinfo  {journal} {Comput. Phys. Commun.}\ }\textbf {\bibinfo
  {volume} {201}},\ \bibinfo {pages} {167--170} (\bibinfo {year} {2016})},\
  \Eprint {http://arxiv.org/abs/1412.5016} {arXiv:1412.5016 [physics.comp-ph]}
  \BibitemShut {NoStop}%
\bibitem [{\citenamefont {Billard}\ \emph {et~al.}(2014)\citenamefont
  {Billard}, \citenamefont {Strigari},\ and\ \citenamefont
  {Figueroa-Feliciano}}]{Billard:2013qya}%
  \BibitemOpen
  \bibfield  {author} {\bibinfo {author} {\bibfnamefont {J.}~\bibnamefont
  {Billard}}, \bibinfo {author} {\bibfnamefont {L.}~\bibnamefont {Strigari}}, \
  and\ \bibinfo {author} {\bibfnamefont {E.}~\bibnamefont
  {Figueroa-Feliciano}},\ }\bibfield  {title} {\enquote {\bibinfo {title}
  {{Implication of neutrino backgrounds on the reach of next generation dark
  matter direct detection experiments}},}\ }\href {\doibase
  10.1103/PhysRevD.89.023524} {\bibfield  {journal} {\bibinfo  {journal} {Phys.
  Rev.}\ }\textbf {\bibinfo {volume} {D89}},\ \bibinfo {pages} {023524}
  (\bibinfo {year} {2014})},\ \Eprint {http://arxiv.org/abs/1307.5458}
  {arXiv:1307.5458 [hep-ph]} \BibitemShut {NoStop}%
\bibitem [{\citenamefont {Ilnicka}\ \emph {et~al.}(2018)\citenamefont
  {Ilnicka}, \citenamefont {Robens},\ and\ \citenamefont
  {Stefaniak}}]{Ilnicka:2018def}%
  \BibitemOpen
  \bibfield  {author} {\bibinfo {author} {\bibfnamefont {A.}~\bibnamefont
  {Ilnicka}}, \bibinfo {author} {\bibfnamefont {T.}~\bibnamefont {Robens}}, \
  and\ \bibinfo {author} {\bibfnamefont {T.}~\bibnamefont {Stefaniak}},\
  }\bibfield  {title} {\enquote {\bibinfo {title} {{Constraining Extended
  Scalar Sectors at the LHC and beyond}},}\ }\href {\doibase
  10.1142/S0217732318300070} {\bibfield  {journal} {\bibinfo  {journal} {Mod.
  Phys. Lett.}\ }\textbf {\bibinfo {volume} {A33}},\ \bibinfo {pages} {1830007}
  (\bibinfo {year} {2018})},\ \Eprint {http://arxiv.org/abs/1803.03594}
  {arXiv:1803.03594 [hep-ph]} \BibitemShut {NoStop}%
\bibitem [{\citenamefont {Fox}\ and\ \citenamefont
  {Poppitz}(2009)}]{Fox:2008kb}%
  \BibitemOpen
  \bibfield  {author} {\bibinfo {author} {\bibfnamefont {P.~J.}\ \bibnamefont
  {Fox}}\ and\ \bibinfo {author} {\bibfnamefont {E.}~\bibnamefont {Poppitz}},\
  }\bibfield  {title} {\enquote {\bibinfo {title} {{Leptophilic Dark
  Matter}},}\ }\href {\doibase 10.1103/PhysRevD.79.083528} {\bibfield
  {journal} {\bibinfo  {journal} {Phys. Rev.}\ }\textbf {\bibinfo {volume}
  {D79}},\ \bibinfo {pages} {083528} (\bibinfo {year} {2009})},\ \Eprint
  {http://arxiv.org/abs/0811.0399} {arXiv:0811.0399 [hep-ph]} \BibitemShut
  {NoStop}%
\bibitem [{\citenamefont {Kopp}\ \emph {et~al.}(2009)\citenamefont {Kopp},
  \citenamefont {Niro}, \citenamefont {Schwetz},\ and\ \citenamefont
  {Zupan}}]{Kopp:2009et}%
  \BibitemOpen
  \bibfield  {author} {\bibinfo {author} {\bibfnamefont {J.}~\bibnamefont
  {Kopp}}, \bibinfo {author} {\bibfnamefont {V.}~\bibnamefont {Niro}}, \bibinfo
  {author} {\bibfnamefont {T.}~\bibnamefont {Schwetz}}, \ and\ \bibinfo
  {author} {\bibfnamefont {J.}~\bibnamefont {Zupan}},\ }\bibfield  {title}
  {\enquote {\bibinfo {title} {{DAMA/LIBRA and leptonically interacting Dark
  Matter}},}\ }\href {\doibase 10.1103/PhysRevD.80.083502} {\bibfield
  {journal} {\bibinfo  {journal} {Phys. Rev.}\ }\textbf {\bibinfo {volume}
  {D80}},\ \bibinfo {pages} {083502} (\bibinfo {year} {2009})},\ \Eprint
  {http://arxiv.org/abs/0907.3159} {arXiv:0907.3159 [hep-ph]} \BibitemShut
  {NoStop}%
\bibitem [{\citenamefont {Bell}\ \emph {et~al.}(2014)\citenamefont {Bell},
  \citenamefont {Cai}, \citenamefont {Leane},\ and\ \citenamefont
  {Medina}}]{Bell:2014tta}%
  \BibitemOpen
  \bibfield  {author} {\bibinfo {author} {\bibfnamefont {N.~F.}\ \bibnamefont
  {Bell}}, \bibinfo {author} {\bibfnamefont {Y.}~\bibnamefont {Cai}}, \bibinfo
  {author} {\bibfnamefont {R.~K.}\ \bibnamefont {Leane}}, \ and\ \bibinfo
  {author} {\bibfnamefont {A.~D.}\ \bibnamefont {Medina}},\ }\bibfield  {title}
  {\enquote {\bibinfo {title} {{Leptophilic dark matter with $Z^\prime$
  interactions}},}\ }\href {\doibase 10.1103/PhysRevD.90.035027} {\bibfield
  {journal} {\bibinfo  {journal} {Phys. Rev.}\ }\textbf {\bibinfo {volume}
  {D90}},\ \bibinfo {pages} {035027} (\bibinfo {year} {2014})},\ \Eprint
  {http://arxiv.org/abs/1407.3001} {arXiv:1407.3001 [hep-ph]} \BibitemShut
  {NoStop}%
\bibitem [{\citenamefont {Freitas}\ and\ \citenamefont
  {Westhoff}(2014)}]{Freitas:2014jla}%
  \BibitemOpen
  \bibfield  {author} {\bibinfo {author} {\bibfnamefont {A.}~\bibnamefont
  {Freitas}}\ and\ \bibinfo {author} {\bibfnamefont {S.}~\bibnamefont
  {Westhoff}},\ }\bibfield  {title} {\enquote {\bibinfo {title} {{Leptophilic
  Dark Matter in Lepton Interactions at LEP and ILC}},}\ }\href {\doibase
  10.1007/JHEP10(2014)116} {\bibfield  {journal} {\bibinfo  {journal} {JHEP}\
  }\textbf {\bibinfo {volume} {10}},\ \bibinfo {pages} {116} (\bibinfo {year}
  {2014})},\ \Eprint {http://arxiv.org/abs/1408.1959} {arXiv:1408.1959
  [hep-ph]} \BibitemShut {NoStop}%
\bibitem [{\citenamefont {del Aguila}\ \emph {et~al.}(2015)\citenamefont {del
  Aguila}, \citenamefont {Chala}, \citenamefont {Santiago},\ and\ \citenamefont
  {Yamamoto}}]{delAguila:2014soa}%
  \BibitemOpen
  \bibfield  {author} {\bibinfo {author} {\bibfnamefont {F.}~\bibnamefont {del
  Aguila}}, \bibinfo {author} {\bibfnamefont {M.}~\bibnamefont {Chala}},
  \bibinfo {author} {\bibfnamefont {J.}~\bibnamefont {Santiago}}, \ and\
  \bibinfo {author} {\bibfnamefont {Y.}~\bibnamefont {Yamamoto}},\ }\bibfield
  {title} {\enquote {\bibinfo {title} {{Collider limits on leptophilic
  interactions}},}\ }\href {\doibase 10.1007/JHEP03(2015)059} {\bibfield
  {journal} {\bibinfo  {journal} {JHEP}\ }\textbf {\bibinfo {volume} {03}},\
  \bibinfo {pages} {059} (\bibinfo {year} {2015})},\ \Eprint
  {http://arxiv.org/abs/1411.7394} {arXiv:1411.7394 [hep-ph]} \BibitemShut
  {NoStop}%
\bibitem [{\citenamefont {Chen}\ \emph {et~al.}(2015)\citenamefont {Chen},
  \citenamefont {Wang},\ and\ \citenamefont {Wang}}]{Chen:2015tia}%
  \BibitemOpen
  \bibfield  {author} {\bibinfo {author} {\bibfnamefont {N.}~\bibnamefont
  {Chen}}, \bibinfo {author} {\bibfnamefont {J.}~\bibnamefont {Wang}}, \ and\
  \bibinfo {author} {\bibfnamefont {X.-P.}\ \bibnamefont {Wang}},\ }\bibfield
  {title} {\enquote {\bibinfo {title} {{The leptophilic dark matter with $Z'$
  interaction: from indirect searches to future $e^+ e^-$ collider
  searches}},}\ }\href@noop {} {\  (\bibinfo {year} {2015})},\ \Eprint
  {http://arxiv.org/abs/1501.04486} {arXiv:1501.04486 [hep-ph]} \BibitemShut
  {NoStop}%
\bibitem [{\citenamefont {D'Eramo}\ \emph {et~al.}(2017)\citenamefont
  {D'Eramo}, \citenamefont {Kavanagh},\ and\ \citenamefont
  {Panci}}]{DEramo:2017zqw}%
  \BibitemOpen
  \bibfield  {author} {\bibinfo {author} {\bibfnamefont {F.}~\bibnamefont
  {D'Eramo}}, \bibinfo {author} {\bibfnamefont {B.~J.}\ \bibnamefont
  {Kavanagh}}, \ and\ \bibinfo {author} {\bibfnamefont {P.}~\bibnamefont
  {Panci}},\ }\bibfield  {title} {\enquote {\bibinfo {title} {{Probing
  Leptophilic Dark Sectors with Hadronic Processes}},}\ }\href {\doibase
  10.1016/j.physletb.2017.05.063} {\bibfield  {journal} {\bibinfo  {journal}
  {Phys. Lett.}\ }\textbf {\bibinfo {volume} {B771}},\ \bibinfo {pages}
  {339--348} (\bibinfo {year} {2017})},\ \Eprint
  {http://arxiv.org/abs/1702.00016} {arXiv:1702.00016 [hep-ph]} \BibitemShut
  {NoStop}%
\bibitem [{\citenamefont {Madge}\ and\ \citenamefont
  {Schwaller}(2019)}]{Madge:2018gfl}%
  \BibitemOpen
  \bibfield  {author} {\bibinfo {author} {\bibfnamefont {E.}~\bibnamefont
  {Madge}}\ and\ \bibinfo {author} {\bibfnamefont {P.}~\bibnamefont
  {Schwaller}},\ }\bibfield  {title} {\enquote {\bibinfo {title} {{Leptophilic
  dark matter from gauged lepton number: Phenomenology and gravitational wave
  signatures}},}\ }\href {\doibase 10.1007/JHEP02(2019)048} {\bibfield
  {journal} {\bibinfo  {journal} {JHEP}\ }\textbf {\bibinfo {volume} {02}},\
  \bibinfo {pages} {048} (\bibinfo {year} {2019})},\ \Eprint
  {http://arxiv.org/abs/1809.09110} {arXiv:1809.09110 [hep-ph]} \BibitemShut
  {NoStop}%
\bibitem [{\citenamefont {Alarcon}\ \emph {et~al.}(2012)\citenamefont
  {Alarcon}, \citenamefont {Martin~Camalich},\ and\ \citenamefont
  {Oller}}]{Alarcon:2011zs}%
  \BibitemOpen
  \bibfield  {author} {\bibinfo {author} {\bibfnamefont {J.~M.}\ \bibnamefont
  {Alarcon}}, \bibinfo {author} {\bibfnamefont {J.}~\bibnamefont
  {Martin~Camalich}}, \ and\ \bibinfo {author} {\bibfnamefont {J.~A.}\
  \bibnamefont {Oller}},\ }\bibfield  {title} {\enquote {\bibinfo {title} {{The
  chiral representation of the $\pi N$ scattering amplitude and the
  pion-nucleon sigma term}},}\ }\href {\doibase 10.1103/PhysRevD.85.051503}
  {\bibfield  {journal} {\bibinfo  {journal} {Phys. Rev.}\ }\textbf {\bibinfo
  {volume} {D85}},\ \bibinfo {pages} {051503} (\bibinfo {year} {2012})},\
  \Eprint {http://arxiv.org/abs/1110.3797} {arXiv:1110.3797 [hep-ph]}
  \BibitemShut {NoStop}%
\bibitem [{\citenamefont {Hoferichter}\ \emph {et~al.}(2017)\citenamefont
  {Hoferichter}, \citenamefont {Klos}, \citenamefont {Menéndez},\ and\
  \citenamefont {Schwenk}}]{Hoferichter:2017olk}%
  \BibitemOpen
  \bibfield  {author} {\bibinfo {author} {\bibfnamefont {M.}~\bibnamefont
  {Hoferichter}}, \bibinfo {author} {\bibfnamefont {P.}~\bibnamefont {Klos}},
  \bibinfo {author} {\bibfnamefont {J.}~\bibnamefont {Menéndez}}, \ and\
  \bibinfo {author} {\bibfnamefont {A.}~\bibnamefont {Schwenk}},\ }\bibfield
  {title} {\enquote {\bibinfo {title} {{Improved limits for Higgs-portal dark
  matter from LHC searches}},}\ }\href {\doibase
  10.1103/PhysRevLett.119.181803} {\bibfield  {journal} {\bibinfo  {journal}
  {Phys. Rev. Lett.}\ }\textbf {\bibinfo {volume} {119}},\ \bibinfo {pages}
  {181803} (\bibinfo {year} {2017})},\ \Eprint
  {http://arxiv.org/abs/1708.02245} {arXiv:1708.02245 [hep-ph]} \BibitemShut
  {NoStop}%
\end{thebibliography}%

\end{document}